\documentclass[fleqn,usenatbib]{mnras}

\usepackage{newtxtext,newtxmath}

\usepackage[T1]{fontenc}
\usepackage{ae,aecompl}

\usepackage{graphicx}	
\usepackage{rotating}
\usepackage{amsmath}	
\usepackage{amssymb}	
\usepackage{multirow}   
\usepackage{longtable}  
\usepackage{threeparttable} 
\usepackage{caption}
\usepackage{array}
\usepackage{ragged2e}
\usepackage{booktabs}       
\usepackage{pdflscape}
\usepackage{siunitx}        
\usepackage{xparse}

\usepackage{url}

\hypersetup{draft}

\ExplSyntaxOn
\NewDocumentCommand{\longdash}{ O{2} }
 {
  --\prg_replicate:nn { #1 - 1 } { \negthinspace -- }
 }
\ExplSyntaxOff

\def\degr{\hbox{$^\circ$}}

\def\arcsec{\hbox{$^{\prime\prime}$}}

\def\gsim{\mathrel{\hbox{\rlap{\lower.55ex \hbox {$\sim$}}
                   \kern-.3em \raise.4ex \hbox{$>$}}}}
\def\lsim{\mathrel{\hbox{\rlap{\lower.55ex \hbox {$\sim$}}
                   \kern-.3em \raise.4ex \hbox{$<$}}}}

\def\he2{\hbox{He\,{\sc ii} $\lambda$4686}}

\def\RL1{\hbox{{$R_{L_{1}}$}}}

\title[Systematic study of spiral density waves]{A systematic study of spiral density waves in the accretion discs of Cataclysmic Variables}

\author[R. Ruiz-Carmona et al.]{
R. Ruiz-Carmona,$^{1}$\thanks{E-mail: r.ruizcarmona@astro.ru.nl}
P. J. Groot,$^{1,2,3,4}$
D. Steeghs$^{5}$
\\
$^{1}$Department of Astrophysics/IMAPP, Radboud University, P.O. Box 9010,
6500 GL Nijmegen, The Netherlands\\
$^{2}$ Department of Astronomy, University of Cape Town, Private Bag X3,
Rondebosch, 7701, South Africa \\
$^{3}$ South African Astronomical Observatory, P.O. Box 9, Observatory,
7935, South Africa \\
$^{4}$ The Inter-University Institute for Data Intensive Astronomy,
University of Cape Town, Private Bag X3, Rondebosch, 7701, South Africa \\
$^{5}$ Department of Physics, University of Warwick, Coventry CV4 7AL, United Kingdom
}

\date{Accepted 2019 October 21. Received 2019 October 15; in original form 2019 July 3}

\pubyear{2019}

\begin{document}
\label{firstpage}
\pagerange{\pageref{firstpage}--\pageref{lastpage}}
\maketitle

\begin{abstract}
Spiral density waves are thought to be excited in the accretion discs of accreting compact objects, including Cataclysmic Variable stars (CVs). Observational evidence has been obtained for a handful of systems in outburst over the last two decades. We present the results of a systematic study searching for spiral density waves in CVs, and report their detection in two of the sixteen observed systems. While most of the systems observed present asymmetric, non-Keplerian accretion discs during outburst, the presence of ordered structures interpreted as spiral density waves is not as ubiquitous as previously anticipated. From a comparison of systems by their system parameters it appears that inclination of the systems may play a major role, favouring the visibility and/or detection of spiral waves in systems seen at high inclination.
\end{abstract}

\begin{keywords}
accretion, accretion discs, shock waves -- binaries: close, spectroscopic -- stars: cataclysmic variables, dwarf novae
\end{keywords}



\section{Introduction}

A wide variety of astrophysical objects, such as newly-formed protostars, compact binaries or supermassive black holes in Active Galactic Nuclei, harbour accretion discs. Matter needs to lose its angular momentum in order to be accreted onto the central object, implying that some form of viscous stress must be operating in accretion discs and driving the transport of angular momentum. \citet{1973A&A....24..337S} suggested that subsonic turbulent eddies, $v_{\rm{turb}} \approx c_s$, comparable in size to the vertical scale height of the disc, $l_{\rm{turb}} \approx H$, were responsible for the viscosity. For the so called $\alpha$-discs, the viscosity can be expressed as: 
\begin{equation}
    \nu = v_{\rm{turb}}\, l_{\rm{turb}} \approx \alpha c_s H,
\end{equation}
where $\alpha$ is a free parameter that characterises the efficiency of the transport of angular momentum.

The viscosity parameter can be constrained by observations, e.g. studies of the evolution of the stellar accretion rate in protostellar discs yield $\alpha \sim 10^{-2}$  \citep{1998ApJ...495..385H}. Observations of compact binaries, and Cataclysmic Variable stars in particular, have provided the best estimates of the $\alpha$ parameter: $\alpha = 0.01 - 0.03$ in cold discs and $\alpha = 0.1 - 0.3$ in hot discs (see reviews by \citealt{1993ApJ...419..318C}, \citealt{2012A&A...545A.115K}). Cataclysmic Variable stars (CVs) consist of a main-sequence late-type star that feeds an accretion disc around a white dwarf via Roche-lobe overflow (see \citealt{1995CAS....28.....W} for an extended review and details of their phenomenology). 

At low mass-transfer rates ($\dot{M} \le 10^{-8}\, M_{\rm{\odot}}\, \rm{yr}^{-1}$), characteristic of the sub-type of non-magnetic CVs known as dwarf novae, accretion discs are susceptible to a thermal-viscous instability occurring when temperatures and densities are high enough to ionise hydrogen (\citealt{1974PASJ...26..429O}, \citealt{1979PThPh..61.1307H}, for a review see \citealt{2001NewAR..45..449L}). This makes their discs alternate between a cold, low accretion-rate, state of quiescence and a hot, high accretion-rate, state of outburst (e.g. \citealt{2012ApJ...757..174C} using U\,Gem and SS\,Cyg as examples). At high mass-transfer rates, characteristic of a subclass of non-magnetic CVs known as novalikes, discs do not experience thermal instability and remain hot at the high state (e.g. \citealt{1993Natur.362..518R} on UX\,UMa, \citealt{2004A&A...417..283G} on RW Tri).

There are several mechanisms proposed to explain the source of viscosity observed in accretion discs (in CVs). First, the magnetorotational instability (MRI; \citealt{1991ApJ...376..214B}) allows conducting plasma cells in differential rotation to exchange angular momentum via magnetic torques induced by a weak axial magnetic field. It has been argued that MRI turbulence is not sufficient to account for the observed viscosity (\citealt{2007MNRAS.376.1740K}, \citealt{2012A&A...545A.115K}), since many numerical simulations produce low values of the viscosity parameter (e.g. \citealt{2010ApJ...708.1716S}, \citealt{2011ApJ...728..130G}), although recent results are trying to reconcile this issue (\citealt{2014ApJ...787....1H}, \citealt{2018ApJ...857...52C}).

\begin{table*}
  \centering
  \renewcommand*{\arraystretch}{1.3}
  \caption{Overview of observations at the William Herschel Telescope}
  \label{tab:runs}
  \begin{tabular}{llll} 
 	
    \hline 
    Programme & Dates & Support Photometric Observations & Comments \\
    \hline
    
    WHT13BN04 & 26--30 Aug 2013 &                  
                                & 8 systems out of 42 candidates\\
 	WHT14AN20 & 19--23 Jun 2014 & Small telescopes, INT, SW14A30     
 	                            & 2 rising outbursts, 3 declining outbursts \\
 	WHT14BN10 & 10--14 Dec 2014 &                               
 	                            & 4 outbursts, SV\,CMi reported in outburst\\
 	WHT15AN10 & 22--26 Jul 2015 & INT, AAVSO Alert Notice 524  
 	                            & 4 outbursts around peak, 2 outbursts declining \\
 	WHT15BN16 & 02--06 Sep 2015 & AAVSO Alert Notice 527       
 	                            & 5 outbursts, V795\,Cyg reported in outburst\\
 	WHT16AN09 & 25--29 May 2016 & AAVSO Alert Notice 543, pt5m 
 	                            & ACAM unavailable. AH\,Her \& RZ\,Sge reported\\ 
 	WHT16BN04 & 05--06 Aug 2016 & & V503\,Cyg reported in outburst \\
 	WHT16BN12 & 16--20 Dec 2016 & AAVSO Alert Notice 564, pt5m 
 	                            & Completely weathered out  \\
 	WHT17AN10 & 11--14 Jul 2017 & AAVSO Alert Notice 586, pt5m 
 	                            & PTFS\,1121b in rising outburst, 4 outbursts around peak\\

 	\hline
  \end{tabular}
\end{table*}

A second contributor to the viscosity are spiral density waves; tidal perturbations that propagate inwards as waves (\citealt{1974MNRAS.168..603L}, \citealt{1986MNRAS.219...75S}, \citealt{1987A&A...184..173S}). Spiral waves can steepen into shocks at speeds faster than the speed of sound in the disc, and deposit negative angular momentum into the disc (\citealt{1980ApJ...241..425G}, \citealt{1984ApJ...285..818P}, \citealt{2018ApJ...854...84A}), especially when the disc is hot and Mach numbers are low, which causes the spiral arms to be more openly wound (\citealt{1993prpl.conf..749L}, \citealt{1999MNRAS.307...99S}). A large number of (magneto-)hydrodynamical (MHD) simulations result in spiral shocks developing in discs (\citealt{1986MNRAS.221..679S}, \citealt{1998MNRAS.297L..81A}, \citealt{1999MNRAS.307...99S}, \citealt{2005ApJ...632..499L}, \citealt{2010MNRAS.405.1397M}), and the first global three-dimensional MHD simulations yield values of the effective viscosity compatible with the observations even if the MRI is operating actively \citep{2016ApJ...823...81J, 2017ApJ...841...29J}. 

Spiral density waves have been directly observed in protoplanetary discs (e.g. \citealt{2016Sci...353.1519P}) and in Saturn's rings by the Cassini mission (e.g. \citealt{2006ApJ...651L..65T}, \citealt{2018Icar..312..157T}). The accretion discs in CVs cannot be spatially resolved, and therefore we rely on indirect imaging techniques, such as Doppler tomography (\citealt{1988MNRAS.235..269M}, see review \citealt{2001LNP...573...B}). Doppler tomography uses the information in the emission line profiles at different orbital phases to construct an image of the accretion disc in velocity space. Spiral density waves have been detected in a handful of systems during outburst, mostly through Doppler tomography (although see \citealt{2001AcA....51..295S} and \citealt{2002MNRAS.330..937O} for alternative interpretations). Most notably IP\,Peg \citep{1997MNRAS.290L..28S}, V347\,Pup \citep{1998MNRAS.299..545S}, EX\,Dra \citep{2000A&A...356L..33J}, U\,Gem \citep{2001ApJ...551L..89G}, WZ\,Sge \citep{2002PASJ...54L...7B} exhibit spiral structure in their tomograms (see \citealt{2001LNP...573...45S} for a review).

In this work, we observe CVs during outburst in order to detect and study spiral density waves with Doppler tomography. We use the hundreds of CVs identified by the Palomar Transient Factory (PTF; \citealt{2009PASP..121.1334R}, \citealt{2009PASP..121.1395L}) to ensure that at any time we can find a number of systems in outburst to follow up spectroscopically. This paper is organised as follows: we describe in detail our observational strategy and describe our observed systems in Section~\ref{sec:obs}; we briefly comment on the analysis techniques and Doppler tomography in Section~\ref{sec:analysis}; we describe our tomographic results in Section~\ref{sec:resuts}, we interpret our results in Section~\ref{sec:discussion} and we summarise our conclusions in Section~\ref{sec:summary}. Additionally, we present photometric data gathered from large synoptic surveys such as PTF in Appendix~\ref{sec:photometry}, we provide an overview of the derivation of system parameters and mass transfer rates in Appendix~\ref{sec:calc_pars}, we briefly comment on the tomograms showing signs of spiral structure published in the literature in Appendix~\ref{sec:comm_sdws}, we present nightly average spectra in Appendix~\ref{sec:avspectra}, and additional tomographic results in Appendix~\ref{sec:dopplermaps}.



\section{Observations}
\label{sec:obs}

\subsection{Observational strategy}
\label{ssec:strategy}

We have used the William Herschel Telescope (WHT) at the Roque de los Muchachos Observatory in La Palma to obtain phase-resolved spectroscopy of CVs in outburst. The occurrence of an outburst in any given system is unpredictable, but we overcame this obstacle thanks to the Palomar Transient Factory (PTF; \citealt{2009PASP..121.1334R}, \citealt{2009PASP..121.1395L}). PTF started observing the optical transient sky in 2009, and it was followed by the intermediate Palomar Transient Factory (iPTF) and more recently by the Zwicky Transient Facility (ZTF; \citealt{2019PASP..131g8001G}, \citealt{2019PASP..131a8002B}). An automated pipeline calibrates all data, performs image subtraction for transient detection and classifies sources in real time \citep{2016PASP..128k4502C}. PTF accumulated the light curves of hundreds of Cataclysmic Variables over the years, so many that it is guaranteed that some of them will be in outburst at any given time. 

PTF and its successors detected and reported transients (and high-amplitude variables such as dwarf nova oubursts) through an internal web-page system called a `transient marshal'. We inspected all the light curves of classified CVs available in the PTF transient marshal, in order to select candidates for our campaigns: as bright as possible and with a recurrence time as short as possible. Light curves with only a few points but exhibiting one or two outbursts will get selected on the basis that it is likely that the system spends a significant fraction of time in outburst, but was located in a field that is sparsely sampled by (i)PTF. This resulted in a number of candidates that were poorly known and for which very little to no additional information is available. Other surveys, such as the Catalina Real-Time Survey (CRTS; \citealt{2009ApJ...696..870D}), were used to confirm the outbursting nature of the systems (see Appendix~\ref{sec:longterm_lc}).

At the beginning of each observing run at the WHT, using the instrument ACAM we performed a quick photometric scan of the candidates. In the first observing run, we found 8 out of 42 candidates in outburst (Table~\ref{tab:runs}). Identification spectra were then acquired on the systems found in outburst in order to select one or two to follow-up with phase-resolved spectroscopy. Selection of the final set of candidates depended on e.g. the brightness, the orbital period, weather conditions and the presence of emission lines during outburst. 

Furthermore, our interest is to study the evolution of the accretion disc as the outburst progresses. In order not to miss the earlier phases of the outburst, we required additional photometric information on the days before our spectroscopic observations. For our second observing programme, we monitored 9 systems with two small telescopes at the Huygens Observatory (Radboud University, Nijmegen, The Netherlands) and at the Rothney Astrophysical Observatory (University of Calgary, Canada). We also obtained photometry on 15 candidates with the Isaac Newton Telescope on 15 June 2014. Additionally, we obtained photometry on 52 systems the night of 18 June 2014 in service mode with WHT/ACAM. With these observations, we found two systems at an early phase of their outburst; most notably PTFS\,1117an for which the rising phase was observed spectroscopically.

In subsequent observing programs, we incorporated the help of amateur astronomers of the American Association of Variable Star Observers\footnote{The American Association of Variable Star Observers (AAVSO) is a non-profit worldwide scientific and educational organization of amateur and professional astronomers who are interested in variable stars. The AAVSO website is \url{https://www.aavso.org}} (AAVSO). Observers received about 20 targets to be monitored two days before our observing runs at the WHT started. A single image per night was enough to confirm the quiescence or outburst state. 

Additional services that report newly found outburst were regularly checked as well, e.g. iPTF Trove Dates (Y. Cao, private communication), the ASASSN CV Patrol \citep{2015AAS...22534402D} and the Cataclysmic Variables Network (CVnet)\footnote{CVnet is a website dedicated to communicate and promote scientific observing campaigns of CVs, latest publications and related news, and collaboration between professional and amateur astronomers. (\url{https://sites.google.com/site/aavsocvsection/Home})} and its \texttt{CVnet-outburst} mailing list service\footnote{\texttt{CVnet-outburst} is a real-time mail notification that reports recent activity of CVs (\url{https://groups.yahoo.com/neo/groups/cvnet-outburst/info})}.

Finally, we also observed some candidates with the robotic telescope {\sl pt5m} (Universities of Durham and Sheffield; \citealt{2015MNRAS.454.4316H}). {\sl pt5m} reduces images and uploads results online in real time, and was very helpful with our photometric scans of candidates. For instance, the system PTFS\,1121b was confirmed in quiescence on 10 July 2017, but found in outburst with pt5m and followed-up with the WHT a day later.

\subsection{Observational data}
\label{ssec:data}

We dedicated 40 nights to our systematic search of spiral density waves with the WHT (see Table~\ref{tab:runs}). We collected around five thousands spectra with the intermediate resolution spectrograph ISIS and several hundred images with ACAM. We present in Table~\ref{tab:observations} details on the observations acquired for every system.

The instrumental set-up for ISIS was to use gratings R600R for the red arm and R600B for the blue arm, resulting in unvignetted wavelength ranges 3950--5100 \AA\; and 5800--7000 \AA\; respectively. The detectors RED+ and EEV12 were set up with a narrow window of 150 pixels in the spatial direction and 2$\times$2 binning, in order to keep the readout time to a minimum and to provide a better match of the pixel scale to the actual resolution of the spectrograph. The slit width was typically 1.0\arcsec, although we have occasionally used wider slits up to 1.5\arcsec\, in poorer seeing conditions. A 1.0\arcsec\, setup gives a velocity dispersion in the range of 22--33 $\rm{km}\, \rm{s}^{-1}$, and an average velocity resolution element of approximately 90 $\rm{km}\, \rm{s}^{-1}$.

All spectra were corrected for the bias level and flatfield variations using an internal lamp flat, and optimally extracted with {\sc iraf} and our own codes in {\sc p}y{\sc raf}. All spectra were acquired in arc-bracketed sequences for wavelength calibration using CuNe and CuAr lamps in, typically, one-hour long runs. The following spectrophotometric standards stars were also observed and used for flux calibration: Feige 66, Feige 67, Feige 110, G191--B2B, GD 50, BD+25 4655, BD+28 4211, BD+33 2642 \citep{1990AJ.....99.1621O}. 

ACAM images were de-biased and corrected with twilight sky flat frames. We performed differential aperture photometry on the targets and 3-4 other field stars, in order to assess the state of ongoing outbursts (see Table~\ref{tab:observations}). The reference magnitudes in the  $r$-filter were obtained from the Panoramic Survey Telescope and Rapid Response System survey (PanSTARRS PS1; \citealt{2010SPIE.7733E..0EK}) and Sloan Digital Sky Survey (SDSS DR15; \citealt{2019ApJS..240...23A}) (see Appendix~\ref{sec:longterm_lc}).

In addition to ACAM images, we obtained photometric observations of PTFS\,1118m with the Wide Field Camera mounted on the 2.5-meter Isaac Newton Telescope on 15 June 2014 (see Appendix~\ref{ssec:lcfitting}).

\subsection{Observational sample}
\label{ssec:sample}

We have observed several subtypes of CVs, characterised by (briefly); SU UMa: orbital periods below the period gap ($<2$ hours) and frequent outbursts (recurrence times are from days to weeks); U\,Gem: orbital periods above the period gap ($>3$ hrs) and regular outbursts (recurrence times are months to years); Z Cam: systems above the period gap that are mostly outbursting like U\,Gem systems but occasionally remain bright in a \lq stand-still\rq; AM CVn: not Cataclysmic Variables, but related systems in which hydrogen-depleted gas is transferred from a white dwarf or semi-degenerate helium-star donor, at orbital periods less than approximately one hour (\citealt{2005ASPC..330...27N}, \citealt{2010PASP..122.1133S}). 

\onecolumn
\begin{center}
\begin{longtable}[c]{lc c c @{\hskip 0.5in}ccc}

 	\caption{Summary of the phase-resolved spectroscopic observations. Included are the results of the aperture photometry performed in the $r$-band images, showing the time of exposure, the magnitude calibrated using PanSTARRS, and the amplitude above the quiescent level.}
 	\label{tab:observations}
 	
 	\\ \hline \hline
    \multirow{2}{1.5cm}{\centering{System}} & 
    \multirow{2}{2.5cm}{\centering{Date}} & 
    \multirow{2}{0.8cm}{$N_{\rm{spec}}$} &         
    \multirow{2}{1.0cm}{$t_{\rm{exp}}$ (s)} &
    \multicolumn{3}{c}{\centering{Photometry}}\\
    
    \cline{5-7}    
    & & & &  UT & $r$ (mag) & $\Delta$ (mag) \\ \hline  \\
    \endhead
        
    \hline \multicolumn{7}{r}{{Continued on next page}} \\ 
    \endfoot

    \hline \hline
    \endlastfoot
        
 	PTFS\,1118m  & 2013-08-26 & 90  & 120 & 21:04 & 15.74 $\pm$ 0.09 & 3.62 $\pm$ 0.11 \\
 		                          & & & & 02:08 & 15.82 $\pm$ 0.07 & 3.55 $\pm$ 0.09 \\
 	           & 2013-08-27 & 120 & 120 & 21:00 & 15.86 $\pm$ 0.03 & 3.51 $\pm$ 0.07 \\
 	                              & & & & 02:20 & 15.79 $\pm$ 0.08 & 3.57 $\pm$ 0.10 \\
 	           & 2013-08-28 & 120 & 120 & 20:37 & 16.02 $\pm$ 0.28 & 3.35 $\pm$ 0.29 \\
 	                              & & & & 01:14 & 16.02 $\pm$ 0.07 & 3.34 $\pm$ 0.09 \\
 	           & 2013-08-29 & 119 & 120 & 20:32 & 16.10 $\pm$ 0.22 & 3.26 $\pm$ 0.23 \\
 	                              & & & & 01:10 & 16.31 $\pm$ 0.08 & 3.06 $\pm$ 0.10 \\
               & 2013-08-30 & 121 & 120 & 20:30 & 17.36 $\pm$ 0.01 & 2.01 $\pm$ 0.07 \\
                                  & & & & 01:10 & 17.61 $\pm$ 0.18 & 1.74 $\pm$ 0.19 \\
 	               
    PTFS\,1122s  & 2013-08-26 & 121 & 60  & 02:27 & 14.17 $\pm$ 0.03 & 5.18 $\pm$ 0.09 \\
 	                              & & & & 05:34 & 14.16 $\pm$ 0.05 & 5.19 $\pm$ 0.10 \\
 	           & 2013-08-27 & 180 & 60  & 02:24 & 14.35 $\pm$ 0.07 & 4.99 $\pm$ 0.11 \\
 	                              & & & & 05:56 & 14.39 $\pm$ 0.03 & 4.95 $\pm$ 0.09 \\
               & 2013-08-28 & 180 & 60  & 01:19 & 14.88 $\pm$ 0.02 & 4.46 $\pm$ 0.09 \\
                                  & & & & 05:47 & 15.02 $\pm$ 0.07 & 4.32 $\pm$ 0.11 \\
 	           & 2013-08-29 & 125 & 120 & 01:18 & 16.02 $\pm$ 0.07 & 3.33 $\pm$ 0.11 \\
 	                              & & & & 05:49 & 16.08 $\pm$ 0.06 & 3.26 $\pm$ 0.11 \\
               & 2013-08-30 &  79 & 180 & 01:20 & 17.07 $\pm$ 0.18 & 2.28 $\pm$ 0.20 \\
                                  & & & & 05:51 & 17.24 $\pm$ 0.07 & 2.11 $\pm$ 0.11 \\
 	               
    PTFS\,1115aa & 2014-06-19 & \longdash  & \longdash  & 23:53 & 15.63 $\pm$ 0.10 & 1.58 $\pm$ 0.10 \\
               & 2014-06-20 & 101 &  90 & 21:09 & 15.89 $\pm$ 0.08 & 1.32 $\pm$ 0.08 \\
                                  & & & & 01:11 & 15.93 $\pm$ 0.03 & 1.28 $\pm$ 0.04 \\
 	           & 2014-06-21 &  43 & 300 & 21:13 & 16.78 $\pm$ 0.03 & 0.43 $\pm$ 0.04 \\
 	                              & & & & 01:16 & 17.04 $\pm$ 0.07 & 0.17 $\pm$ 0.14 \\
 	           & 2014-06-22 &  42 & 300 & 01:08 & 17.39 $\pm$ 0.06 & --0.18 $\pm$ 0.14 \\
               & 2014-06-23 &  42 & 300 & 21:07 & 17.38 $\pm$ 0.10 & --0.17 $\pm$ 0.16 \\
 	                              & & & & 01:12 & 17.51 $\pm$ 0.04 & --0.30 $\pm$ 0.13 \\
    
    PTFS\,1117an & 2014-06-20 &  63 & 180 & 21.33 & 16.16 $\pm$ 0.03 & 0.23 $\pm$ 0.14 \\
 	                              & & & & 01:17 & 16.00 $\pm$ 0.01 & 0.38 $\pm$ 0.13 \\
 	           & 2014-06-21 &  61 & 180 & 01:29 & 14.73 $\pm$ 0.05 & 1.65 $\pm$ 0.14 \\
 	                              & & & & 05:10 & 14.33 $\pm$ 0.03 & 2.05 $\pm$ 0.13 \\
 	           & 2014-06-22 &  80 & 150 & 01:20 & 15.29 $\pm$ 0.07 & 1.10 $\pm$ 0.15 \\
 	                              & & & & 05:24 & 15.40 $\pm$ 0.07 & 0.98 $\pm$ 0.15 \\
 	           & 2014-06-23 &  83 & 150 & 01:22 & 15.80 $\pm$ 0.03 & 0.58 $\pm$ 0.13 \\
 	                              & & & & 05:19 & 16.09 $\pm$ 0.15 & 0.30 $\pm$ 0.20 \\
 	
 	PTFS\,1200o  & 2014-12-11 & \longdash  & \longdash  & 19:38 & 16.59 $\pm$ 0.03 & 2.13 $\pm$ 0.05 \\  
 	           & 2014-12-12 &  30 & 450 & 19:10 & 16.73 $\pm$ 0.10 & 2.00 $\pm$ 0.14 \\  
 	           & 2014-12-13 &  37 & 550 & 19:41 & 16.63 $\pm$ 0.10 & 2.10 $\pm$ 0.14 \\
 	           & 2014-12-14 &  29 & 550 & 20:47 & 16.73 $\pm$ 0.04 & 1.99 $\pm$ 0.05 \\    
 	                        
 	SV\,CMi     & 2014-12-11 & \longdash & \longdash & 05:52 & 15.71 $\pm$ 0.14 & 1.48 $\pm$ 0.15 \\    
 	           & 2014-12-12 &   2 & 550 & 01:10 & 16.00 $\pm$ 0.10 & 1.19 $\pm$ 0.12 \\    
 	           & 2014-12-13 &  19 & 360 & 02:30 & 16.36 $\pm$ 0.10 & 0.83 $\pm$ 0.11 \\   
 	           & 2014-12-14 &  29 & 360 &  \\  
 	           
    PTFS\,1119h  & 2015-07-22 & 142 &  90 & 21:46 & 14.20 $\pm$ 0.09 & 3.14 $\pm$ 0.23 \\
 	                              & & & & 02:47 & 14.27 $\pm$ 0.08 & 3.07 $\pm$ 0.23 \\
 	           & 2015-07-23 & 163 &  90 & 20:52 & 14.31 $\pm$ 0.16 & 3.03 $\pm$ 0.27 \\    
 	           & 2015-07-24 & 181 &  90 & 20:49 & 14.66 $\pm$ 0.13 & 2.68 $\pm$ 0.25 \\  
 	                              & & & & 02:36 & 14.60 $\pm$ 0.03 & 2.74 $\pm$ 0.22 \\
 	           & 2015-07-25 & 160 &  90 & 20:48 & 14.86 $\pm$ 0.14 & 2.48 $\pm$ 0.26 \\    
 	                              & & & & 02:06 & 14.58 $\pm$ 0.04 & 2.76 $\pm$ 0.22 \\  
 	           & 2015-07-26 & 120 &  90 & 22:26 & 14.66 $\pm$ 0.12 & 2.68 $\pm$ 0.25 \\       
    \pagebreak      
    PTFS\,1101af & 2015-09-02 & \longdash & \longdash & 05:43 & 14.02 $\pm$ 0.04 & 2.01 $\pm$ 0.13 \\   	
               & 2015-09-03 &  84 & 150 & 01:56 & 14.33 $\pm$ 0.07 & 1.70 $\pm$ 0.14 \\
 	                              & & & & 06:04 & 14.31 $\pm$ 0.08 & 1.72 $\pm$ 0.15 \\
               & 2015-09-04 &  88 & 150 & 01:58 & 15.08 $\pm$ 0.13 & 0.95 $\pm$ 0.18 \\   
 	                              & & & & 06:09 & 15.32 $\pm$ 0.02 & 0.71 $\pm$ 0.12 \\ 
         	   & 2015-09-05 &  90 & 150 & 01:54 & 15.73 $\pm$ 0.03 & 0.30 $\pm$ 0.12 \\ 
         	   & 2015-09-06 &  86 & 150 & 02:11 & 16.09 $\pm$ 0.04 & --0.06 $\pm$ 0.13 \\ 
 	                        
    V795\,Cyg   & 2015-09-02 &  91 & 180 & 02:28 & 15.08 $\pm$ 0.74 & 3.31 $\pm$ 0.75 \\ 
               & 2015-09-03 & 170 &  90 & 20:38 & 14.13 $\pm$ 0.68 & 4.26 $\pm$ 0.68 \\ 
               & 2015-09-04 & 170 &  90 & 01:49 & 14.08 $\pm$ 0.39 & 4.32 $\pm$ 0.39 \\
         	   & 2015-09-05 & 151 &  90 & 20:29 & 14.12 $\pm$ 0.41 & 4.28 $\pm$ 0.41 \\   
         	   & 2015-09-06 & 190 &  90 & 20:36 & 15.73 $\pm$ 0.55 & 2.66 $\pm$ 0.55 \\ 
 	                        
    AH\,Her     & 2016-05-27 &  72 & 240 & 21:39 & 12.21 $\pm$ 0.03 & 2.44 $\pm$ 0.03 \\  
 	                              & & & & 05:06 & 12.11 $\pm$ 0.13 & 2.54 $\pm$ 0.13 \\    
    
    PTFS\,1113bc & 2016-05-27 & \longdash & \longdash & 21:27 & 17.07 $\pm$ 0.02 & --0.27 $\pm$ 0.13 \\
               & 2016-05-28 &  62 &  60 & 21:43 & 16.22 $\pm$ 0.14 &  0.58 $\pm$ 0.19 \\  
 	                              & & & & 23:07 & 16.11 $\pm$ 0.10 &  0.69 $\pm$ 0.16 \\    
               & 2016-05-29 & \longdash & \longdash & 21:33 & 16.98 $\pm$ 0.04 & --0.18 $\pm$ 0.14 \\
               
    PTFS\,1215t  & 2016-05-27 & \longdash & \longdash & 21:34 & 16.34 $\pm$ 0.04 & 2.53 $\pm$ 0.11 \\
               & 2016-05-28 &  67 & 120 & 23:11 & 16.89 $\pm$ 0.06 & 1.98 $\pm$ 0.12 \\  
               & 2016-05-29 &  85 & 180 & 21:50 & 17.93 $\pm$ 0.03 & 0.94 $\pm$ 0.11 \\  
    
    RZ\,Sge     & 2016-05-28 & 107 &  60 & 02:55 & 13.05 $\pm$ 0.19 & 4.05 $\pm$ 0.19 \\  
               & 2016-05-29 & 120 &  60 & 02:58 & 12.93 $\pm$ 0.08 & 4.17 $\pm$ 0.09 \\  
 	                        
    V503\,Cyg   & 2016-08-05 &  80 & 120 & \\  
 	        
 	PTFS\,1121b  & 2017-07-11 &  45 & 180 & 01:49 & 13.33 $\pm$ 0.09 & 4.53 $\pm$ 0.09 \\
 	                              & & & & 05:11 & 13.65 $\pm$ 0.15 & 4.21 $\pm$ 0.15 \\ 
 	           & 2017-07-12 &  65 & 180 & 01:38 & 13.70 $\pm$ 0.11 & 4.16 $\pm$ 0.11 \\
 	           & 2017-07-13 &  75 & 180 & 01:10 & 13.57 $\pm$ 0.08 & 4.29 $\pm$ 0.08 \\
 	           & 2017-07-14 &  47 & 300 & 01:16 & 13.93 $\pm$ 0.06 & 3.92 $\pm$ 0.06 \\  
 	                              & & & & 05:36 & 14.13 $\pm$ 0.08 & 3.74 $\pm$ 0.08 \\
 	                        
 	PTFS\,1716bc & 2017-07-11 &  28 & 360 & 22:08 & 17.63 $\pm$ 0.08 & 1.50 $\pm$ 0.15 \\  
 	                              & & & & 01:22 & 18.15 $\pm$ 0.04 & 0.98 $\pm$ 0.13 \\ 
 	           & 2017-07-12 &  28 & 360 & 21:34 & 18.03 $\pm$ 0.10 & 1.10 $\pm$ 0.16 \\   
 	                              & & & & 00:58 & 18.14 $\pm$ 0.20 & 0.99 $\pm$ 0.24 \\
 	           & 2017-07-13 &  26 & 400 & 21:27 & 18.02 $\pm$ 0.13 & 1.11 $\pm$ 0.18 \\  
 	                              & & & & 00:52 & 18.05 $\pm$ 0.35 & 1.07 $\pm$ 0.37 \\
 	           & 2017-07-14 &  22 & 450 & 01:10 & 18.71 $\pm$ 0.36 & 0.42 $\pm$ 0.38 \\  
 	           
 		\hline
 		
\end{longtable}
\end{center}
\twocolumn

\begin{table*}
  \centering
  \renewcommand*{\arraystretch}{1.3}
  \caption{Classification and coordinates of the observed systems.}
  \label{tab:coords}
  \begin{tabular}{llccc} 
 	
    \hline 
    \multirow{2}{1.2cm}{\centering{System}} & 
    \multirow{2}{3cm}{Alternative names} &
    \multirow{2}{0.8cm}{\centering{Class}} & 
    RA & DEC \\
    & & & (h m s) & (d m s) \\
    \hline
    
    PTFS\,1118m  & ASASSN-13bc & SU UMa & 18 02 22.44 & +45 52 44.5 \\
    PTFS\,1122s  & V650 Peg    & SU UMa & 22 43 48.48 & +08 09 26.9 \\
 	PTFS\,1115aa & NY Ser      & SU UMa & 15 13 02.30 & +23 15 08.5 \\
 	PTFS\,1117an & SDSS J173008.38+624754.7 & SU UMa & 17 30 08.36 & +62 47 54.6 \\
 	PTFS\,1200o  & SDSS J005347.32+405548.5 & U\,Gem  & 00 53 47.33 & +40 55 48.6 \\
 	SV\,CMi     &             & Z Cam  & 07 31 08.41 & +06 05 11.2 \\
 	PTFS\,1119h  & V1504 Cyg   & U\,Gem  & 19 28 56.45 & +43 05 36.4 \\ 
 	PTFS\,1101af & AR And      & U\,Gem  & 01 45 03.27 & +37 56 33.3 \\
 	V795\,Cyg   & PTFS\,1719e   & U\,Gem  & 19 34 34.32 & +31 32 11.1 \\
 	AH\,Her     & PTFS\,1516bg  & Z Cam  & 16 44 10.01 & +25 15 01.9 \\
 	PTFS\,1113bc & CR Boo      & AM CVn & 13 48 55.19 & +07 57 35.9 \\
 	PTFS\,1215t  & SDSS J153015.04+094946.3 & SU UMa & 15 30 15.03 & +09 49 46.5 \\
 	RZ\,Sge     & PTFS\,1720h   & SU UMa & 20 03 18.55 & +17 02 52.3 \\
 	V503\,Cyg   & PTFS\,1720i   & SU UMa & 20 27 17.41 & +43 41 22.5 \\
 	PTFS\,1121b  & VZ Aqr      & U\,Gem  & 21 30 24.61 & --02 59 17.6 \\
 	PTFS\,1716bc &             & SU UMa & 16 19 53.50 & +03 19 09.6 \\
 	
 	\hline
  \end{tabular}
\end{table*}

An overview of all observed systems is given in Table~\ref{tab:coords}. Our sample contains 8 systems of the SU UMa sub-type, 6 systems of the U\,Gem sub-type, 1 system of the Z Cam sub-type and 1 AM CVn system. The long-term light curves of the stars in our sample, as observed by different surveys, are presented in Figures~\ref{fig:monitoring1}-\ref{fig:monitoring2}. We have also calculated their system parameters (see Appendix~\ref{sec:calc_pars} for details), and present the results in Table~\ref{tab:ours_pars}.

A number of systems are noteworthy: 

\begin{itemize}

    \item PTFS\,1118m is a relatively unknown system, that started being followed by amateur observers after ASASSN reported the first outburst \citep{vsnet-15906}. We report its orbital period obtained via light curve fitting (see Appendix~\ref{ssec:lcfitting}). 

    \item PTFS\,1115aa (NY Ser) is a SU UMa-subtype star in the period gap, that shows stand-stills typical of Z Cam stars \citep{2019PASJ...71L...1K}. The observations of PTFS\,1115aa by \citet{2019PASJ...71L...1K} provided the first evidence that superoutbursts can begin from standstills, and that accretion discs can grow during standstill state.

    \item PTFS\,1117an, also SDSS\,J173008.38+624754.7, has been well observed by several surveys (\citealt{2009ApJ...696..870D}, \citealt{2013AJ....146..101P}; see Figure~\ref{fig:monitoring1}) and has been included in a handful of studies of statistical properties of CVs (e.g. \citealt{2016MNRAS.460.2526O}).

    \item PTFS\,1200o is an eclipsing system extensively covered by PTF and CRTS. Its light curve exhibits primary and secondary eclipses, that were initially confused with one another in the PTF transient marshal. In this work, we model the light curve and report its orbital period (see Section~\ref{ssec:lcfitting}).

    \item SV\,CMi is a bright U\,Gem subtype CV discovered in the late 1920s (\citealt{1929MiSon..16....1H}, \citealt{1930AN....238...17H}), that was mis-classified as Z Cam \citep{2014JAVSO..42..177S}. There are a large number of observations in the AAVSO database, but the system has not been observed by any large synoptic survey, presumably because of its proximity to the Galactic Plane ($b=12.1\degr$). SV\,CMi was reported in outburst by \texttt{CVnet-outburst} services on 11 December 2014.

    \item PTFS\,1119h (V1504 Cyg) is a SU UMa star that was extensively observed by \textit{Kepler}. \citet{2013PASJ...65...50O} studied the \textit{Kepler} light curve and recognized that every single superoutburst began with a precursor normal outburst, with superhumps appearing at the peak of the precursor outburst. \citet{2013PASJ...65...50O} also showed that positive and negative superhumps can coexist in this system, which means that the disc can be eccentric and tilted at the same time. The frequency of negative superhumps varies systematically with the supercycle, providing evidence of disc growth in radius which supports the thermal-tidal instability model \citep{1989PASJ...41.1005O}. We first observed PTFS\,1119h in our June 2014 observing run, but we failed to cover the entire orbital period. We obtained spectroscopic data in July 2015, when the system was probably in superoutburst; the system declined 0.46 mag in 5 days according to our photometric data (Table~\ref{tab:observations}).

    \item PTFS\,1101af (AR And) is a bright system that goes into outburst very frequently \citep{2016MNRAS.456.4441C}. \citet{1995ApJ...440..853S} presented a Doppler tomogram of H$\alpha$ in emission, where the disc appears axisymmetric with no additional features. \citet{1996PASP..108..894T} obtained phase-resolved spectroscopy in order to obtain the orbital period. Their average spectrum in quiescence shows narrow emission lines, with relatively strong Fe {\sc ii} lines that fall outside of the wavelength range of our observations.

    \item AH\,Her is a system of the Z Cam subtype which has been very well observed by AAVSO amateur observers \citep{2014JAVSO..42..177S}, and other surveys (e.g. \citealt{2009ApJ...696..870D}) thanks to its intrinsic brightness in quiescence. AH\,Her's spectrum is dominated by the accretion disc and the secondary star, with minimal contribution of the white dwarf (\citealt{2006ApJ...642.1029U}, \citealt{2007ApJ...667.1139H}). Intense photometric monitoring of AH\,Her has been carried out since 1994 (e.g. \citealt{2001IBVS.5147....1S}, \citealt{2002IBVS.5276....1S}). A set of synthetic disc spectra at different levels of activity has been computed \citep{2011IBVS.5983....1S}.

    \item CR Boo (PTFS\,1113bc) is one of the best known AM CVn stars. The system's long-term behaviour alternates between a state of frequent outbursts, spending 74\% of the time at around peak luminosity \citep{1987ApJ...313..757W} and a state of regular superoutbursts with normal precursors with a fainter quiescence \citep{2016PASJ...68...64I}. Its light curve shows a very rapid decay from outburst, i.e. $2.7\, \rm{mag}\, \rm{d}^{-1}$ \citep{2000MNRAS.315..140K}, and its spectrum shows wide and shallow neutral helium lines \citep{1987ApJ...313..757W}. \citet{2015MNRAS.446..391L} showed the system to have a 46.5d recurrence time between superoutbursts. We obtained phase-resolved spectroscopy on 28 May 2016, as the system seemed to be active on that night but in quiescence state the previous and the following nights.

    \item PTFS\,1215t (SDSS\,J153015.04+094946.3), is a known SU UMa subtype star characterised by a short supercycle of approximately 85 days, but unexpectedly only shows a small number of normal outbursts \citep{2017PASJ...69...75K}.  \citet{2007ApJ...661.1042W} proposed that the accretion disc must be warped or tilted, allowing for the accreted matter to reach the inner parts of the disc instead of accumulating at the outer edge. 

    \item RZ\,Sge is a SU UMa subtype star, well monitored by amateur observers. Its spectrum presents double-peaked H$\alpha$ line \citep{2003PASP..115.1308P}, indicative of high inclination. 

    \item V503\,Cyg is known to exhibit both positive and negative superhumps \citep{1995PASP..107..551H}, although the negative superhumps are not always present \citep{2012Ap.....55..494P}. V503\,Cyg presents states of very frequent normal outbursts in the supercycle and also states of very normal outbursts accompanied by premature quenching of superoutbursts \citep{2002PASJ...54.1029K}. Similarly to PTFS\,1119h and PTFS\,1215t, a tilted disc is invoked in order to explain the later state. V503\,Cyg was reported into outburst by the \texttt{CVnet-outburst} services, and we lack photometric observations.

    \item PTFS\,1121b (VZ Aqr) is a very little-studied dwarf nova, despite being a relatively bright system during outburst. Its recurrence time has been derived \citep{1994PZ.....23..226S}, and its outburst spectrum shows absorption lines with emission cores for the Balmer lines, with very faint helium lines \citep{2002MNRAS.332..814M}.

    \item PTFS\,1716bc is a dwarf nova discovered by \citet{2006A&A...449...79R} in a search of optical afterglows of gamma-ray bursts. We found that PTFS\,1716bc is an eclipsing system, and report its orbital period (Section~\ref{ssec:lcfitting}).

\end{itemize}


\section{Analysis}
\label{sec:analysis}

All spectra were normalised by their continuum levels, that were fitted with a low-order polynomial. Spectra were rebinned in velocity space around the central wavelength of the lines, and saved in the {\sc ndf} format used by the {\sc molly}\footnote{{\sc molly} and {\sc doppler} are a software package to analyze spectra and perform Doppler tomography, courtesy of Tom Marsh.} software. 

Since Doppler tomography, in the maximum-entropy implementation of {\sc doppler}, cannot deal with negative numbers, we treated the absorption lines present in some of the spectra (see Figures~\ref{fig:avspec1}-\ref{fig:avspec2}) in the following manner, following \citet{1990ApJ...364..637M}. We fitted the absorption on the average spectrum per night with a Gaussian profile, masking out the central cores in emission. We built a two-dimensional Gaussian in Doppler space with the width and peak flux obtained in the fit. We derived spectra from the 2D Gaussian at identical pixel size, phases and exposures times to our data, and we added them to the original spectra. This is sufficient to negate the absorption and recover the emission cores. Note that adding a 2D Gaussian will not contribute to any asymmetries that may appear in the tomograms.

We computed Doppler tomograms with the {\sc doppler}\footnote{{\sc doppler} is a free software that compute Doppler maps, also courtesy of Tom Marsh.} software. We implemented a proportional decrement in $\chi^2$, and we blurred the tomograms with a Gaussian that is roughly 4 resolution elements wide at every step. We derived systemic velocities for every data set, using radial velocity analysis, diagnostic diagrams and Doppler tomography itself (see \citealt{doppler_methods}). These techniques yielded similar results for the strongest lines, but results by Doppler tomography were favoured if there is disagreement for the faintest lines. We selected the optimal tomograms with the guidance of two-dimensional Fourier transforms of the tomograms \citep{doppler_methods}. In cases where the absolute phase is unknown, we added an offset to locate the presumed signature of the secondary star at the expected position in the tomograms. Some of the data sets are too faint for Doppler tomograms to be computed; trailed spectra only are presented instead.

We apply a bootstrap routine to a selection of systems that exhibit spiral structure in their tomograms in order to obtain significance maps following \citet{doppler_methods}. In brief, we compute $N_{\rm{boots}} = 2500$ tomograms from data subsets for which pixels are randomly selected with replacement. We adjusted error bars according to the number of times a particular pixel was selected, and conserve the flux \citep{2001MNRAS.326...67W}. We define a representative region of the accretion disc, typically towards the lower right region of the tomograms (orbital phases 0.25--0.5), where no emitting features such as spiral waves or bright spots are expected. The radial origin of these regions is at the position of the white dwarf, and we use constraints in flux to delimit the upper and lower edge in (velocity) radius from the white dwarf; and add a constraint in azimuth to avoid including emitting structures. We characterise the flux in the accretion disc region by its average, $F_{\rm{disc}}$, and its standard deviation $\sigma_{\rm{disc}}$. We obtain significance maps on a pixel by pixel basis, by taking the largest value of $m$ that verifies the null hypothesis that the flux in a pixel is larger than the flux in the selected disc region plus $m$ times the standard deviation of the disc: 
\begin{equation}
    H_0: F_{\rm{pix}} > F_{\rm{disc}} + m \sigma_{\rm{disc}}
\end{equation}

We emphasize that the variance parameter $m$ cannot be interpreted in terms of $\sigma$ significance, because the parent distribution is not necessarily Gaussian. It is a lower limit since the standard deviation in the accretion disc is always larger than the standard deviation of the distribution of flux of any pixel across the bootstrapped tomograms, $\sigma_{\rm{pix}}$:  $\sigma_{\rm{disc}} \gsim \sigma_{\rm{pix}}$.


\section{Results}
\label{sec:resuts}

For all our systems, we present an average spectrum per observing night in Figures~\ref{fig:avspec1}-\ref{fig:avspec2}, and trailed spectra and Doppler tomograms of a selection of the strongest spectral lines in Figures~\ref{fig:tom_ptfs1118m}-\ref{fig:bootstraps}.

We classify the resulting tomograms in three groups:
\begin{enumerate}[i]
    \item Doppler maps that show very little structure. The accretion disc is difficult to distinguish. Some of these tomograms are dominated by small regions of enhanced emission, that cannot be traced to known or expected components such as bright spots or spiral shocks.
    
    \item Tomograms characterised by asymmetric, non-Keplerian discs, with emission components that appear at regions where spiral shocks are not expected. 
    
    \item Tomograms with evidence of spiral structure. We further examine these maps using bootstrapping in order to confirm the significance of the structures.
\end{enumerate}    
   
\vspace{-0.2cm}   
\subsection{Tomograms with undefined discs}
\label{ssec:res-nodisc}

\begin{figure*}
  \caption{Doppler tomograms and trailed spectra for PTFS\,1118m}
  \label{fig:tom_ptfs1118m}
  \centering
    \includegraphics[width=0.89\textwidth]{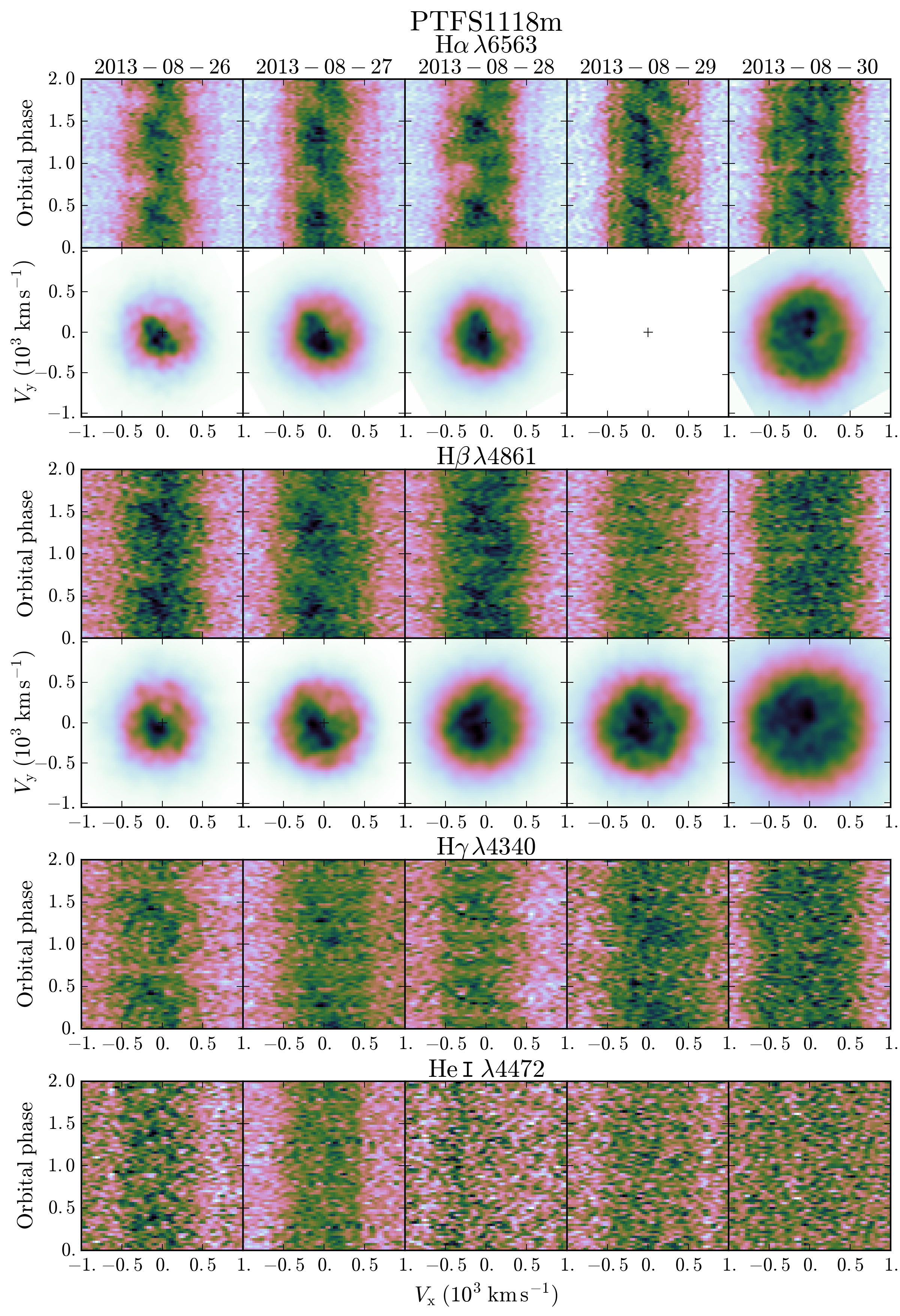}
\end{figure*}

\begin{figure*}
  \caption{Doppler tomograms and trailed spectra for PTFS\,1122s}
  \label{fig:tom_ptfs1122s}
  \centering
    \includegraphics[width=0.89\textwidth]{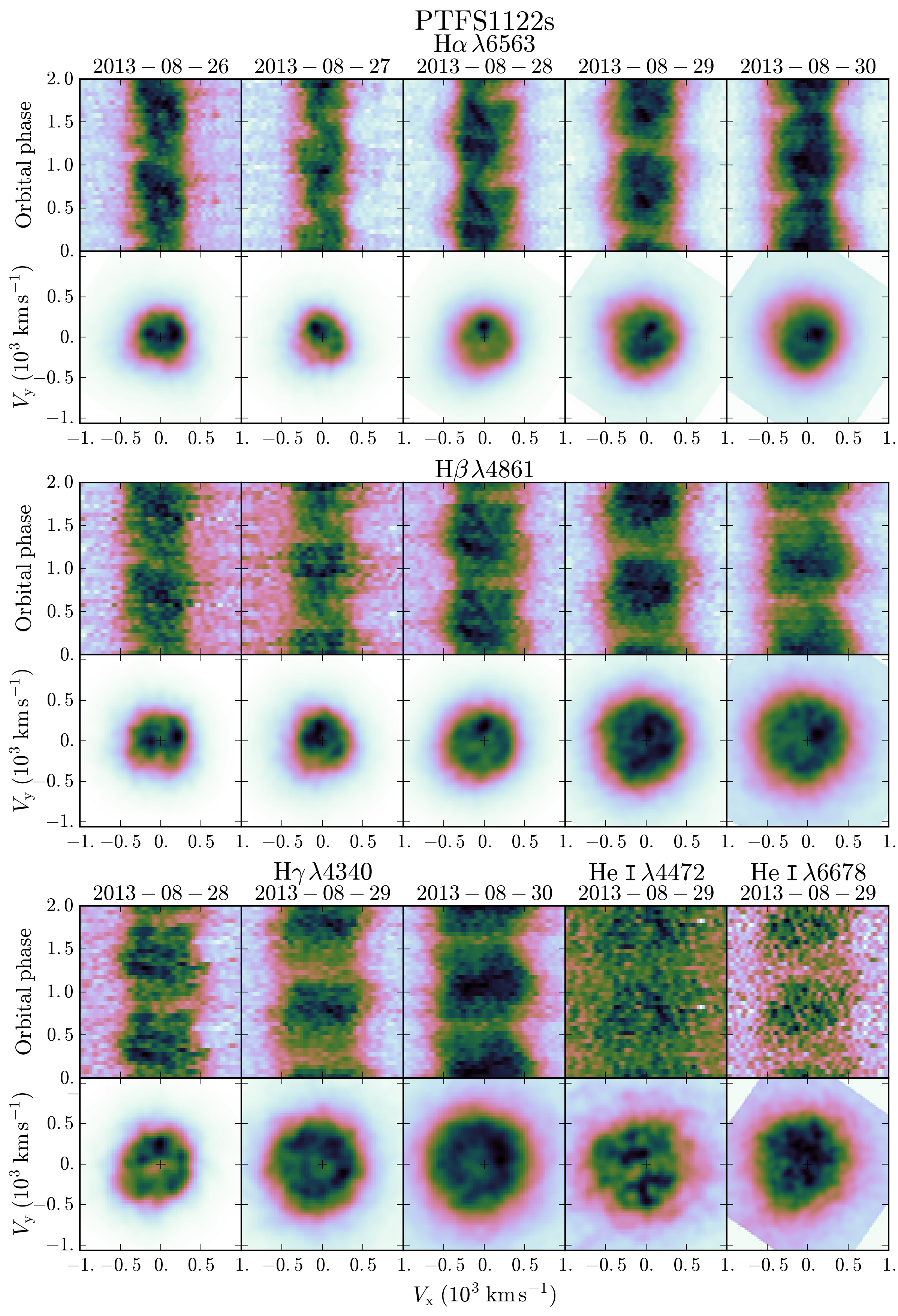}
\end{figure*}

\begin{figure*}
  \caption{Doppler tomograms and trailed spectra for PTFS\,1119h}
  \label{fig:tom_ptfs1119h}
  \centering
    \includegraphics[width=0.89\textwidth]{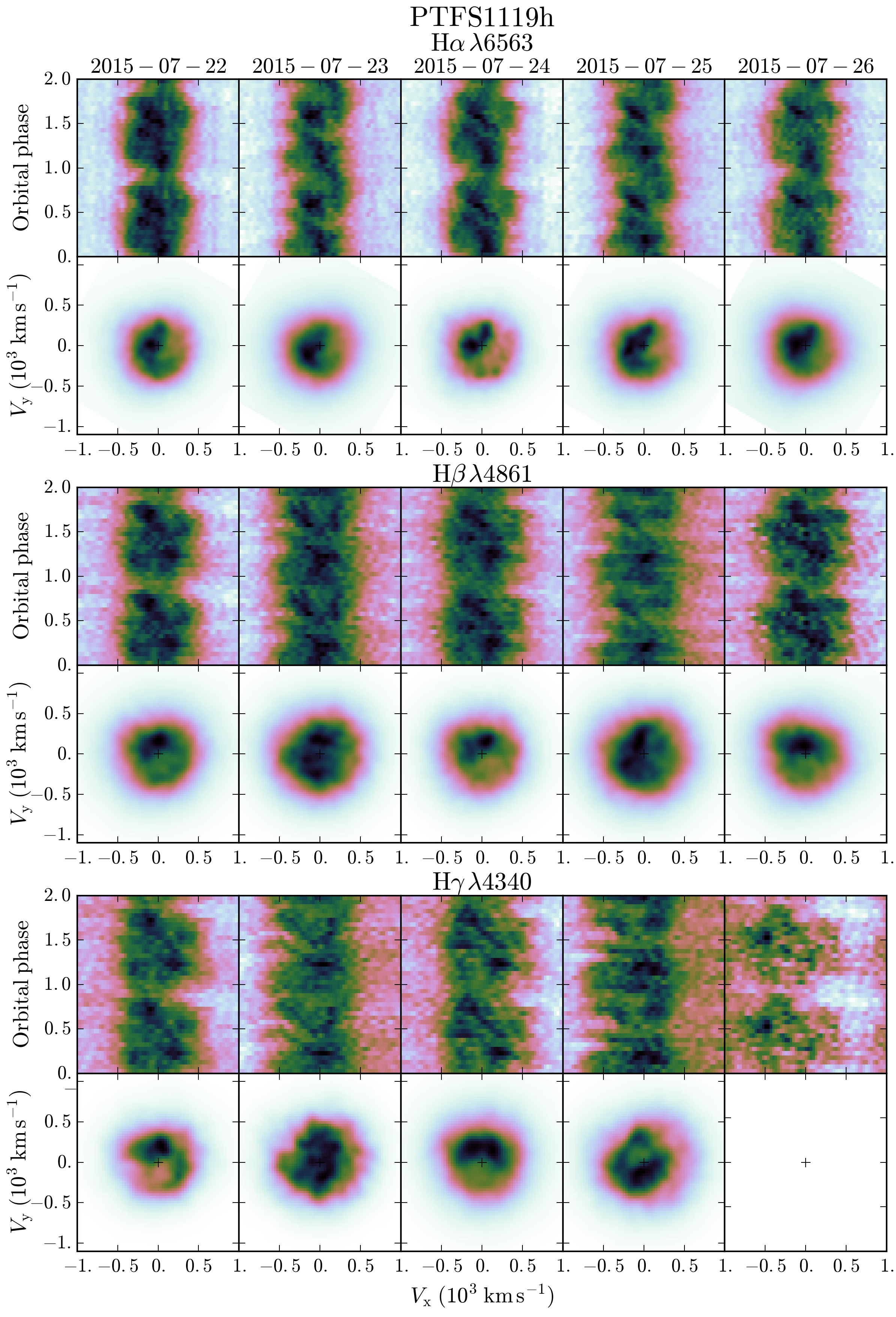}
\end{figure*}

\begin{landscape}
\begin{figure}
  \caption{Doppler tomograms and trailed spectra for PTFS\,1215t, PTFS\,1113bc (CR Boo) and V503\,Cyg}
  \label{fig:tom_ptfs1215t-CR Boo-v503cyg}
  \centering
    \includegraphics[width=1.35\textheight]{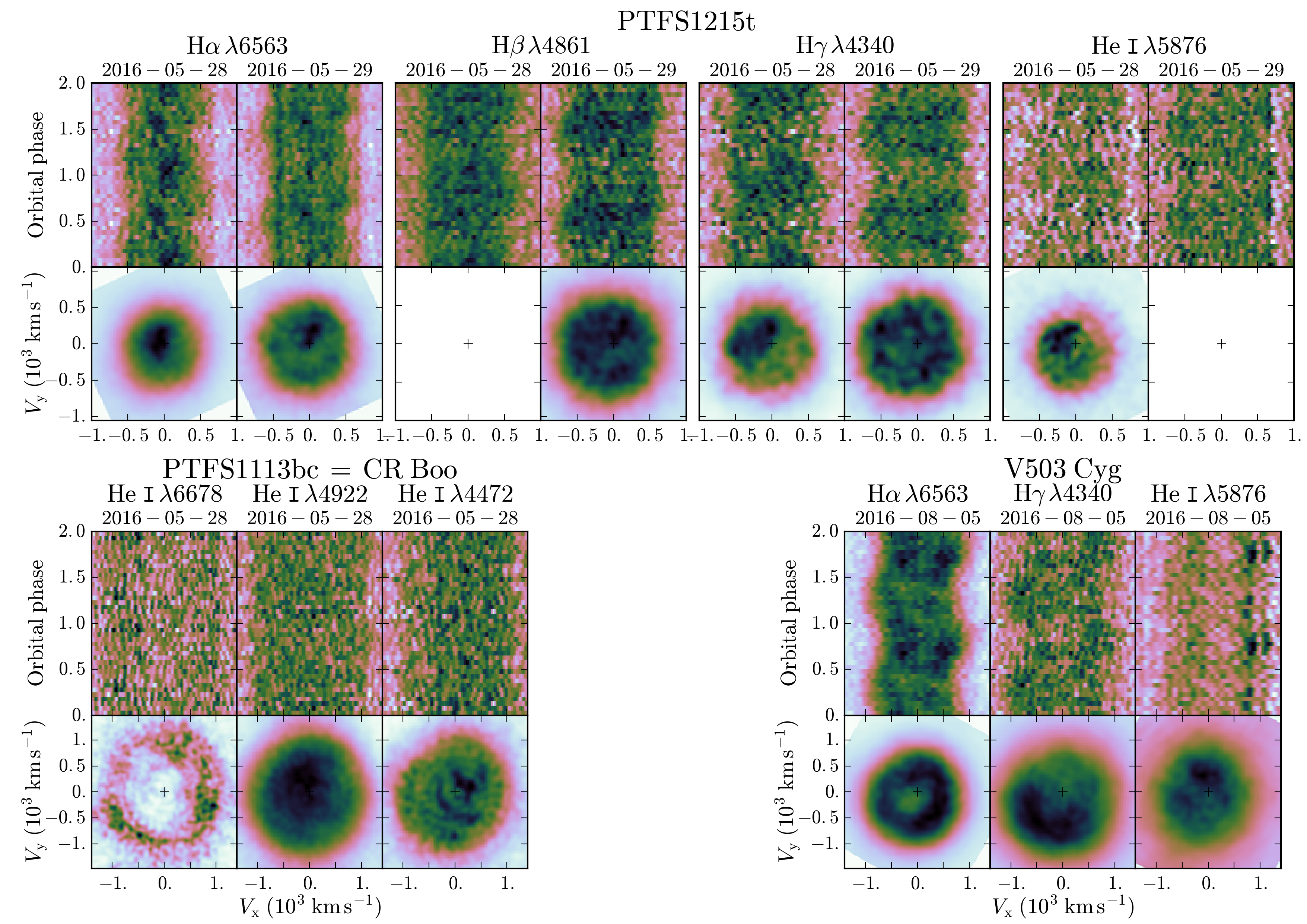}
\end{figure}
\end{landscape}

The systems showing little to no structure in their tomograms are:

\begin{itemize}
    \item PTFS\,1118m (Figure~\ref{fig:tom_ptfs1118m}); an undefined structure close to the center of the tomograms is visible in the Balmer lines. The secondary star and a central spike are clearly visible in H$\alpha$ on August 30. 
        
    \item PTFS\,1122s (Figure~\ref{fig:tom_ptfs1122s}); tomograms are rotated in phase assuming that the secondary star is observed in H$\alpha$ on August 28. This leaves the presumed emission from the secondary at different nights out of phase, questioning our assumption. Tomograms of the Balmer lines have a consistent appearance from night to night, exhibiting some structure in the disc towards the end of our observations. The most consistent emission feature appears towards the lower right part of the maps. The tomograms of He {\sc i} lines are extremely faint. We also note consistent variations in the equivalent width of the lines showing up as horizontal bars in the trailed spectra, most notably on August 29, perhaps hinting at self-obscuration.
        
    \item PTFS\,1119h (Figure~\ref{fig:tom_ptfs1119h}); S-waves in the trailed spectra translate into extended emission in the tomograms. This emission component is consistent across the Balmer lines for a given night, but changes in extension and phase with time. It is hard to recognize the accretion disc or emission from the secondary star.
        
    \item PTFS\,1215t (Figure~\ref{fig:tom_ptfs1215t-CR Boo-v503cyg}); does not show a clear accretion disc nor any other distinguishable structure, possibly due to lower signal-to-noise ratios.
        
    \item PTFS\,1113bc (CR Boo; Figure~\ref{fig:tom_ptfs1215t-CR Boo-v503cyg}): the spectral lines are extremely faint and absorption is hard to model. As a result, the tomograms are very faint.
        
    \item V503\,Cyg (Figure~\ref{fig:tom_ptfs1215t-CR Boo-v503cyg}); the accretion disc in H$\alpha$ shows enhanced emission towards the right part of the tomogram. In H$\gamma$, the enhanced emission appears at the lower left of the tomogram, and occupies half of the disc.
    
\end{itemize}

\subsection{Tomograms with asymmetric discs}
\label{ssec:res-asym}

\begin{landscape}
\begin{figure}
  \caption{Doppler tomograms and trailed spectra for PTFS\,1115aa}
  \label{fig:tom_ptfs1115aa}
  \centering
    \includegraphics[width=1.35\textheight]{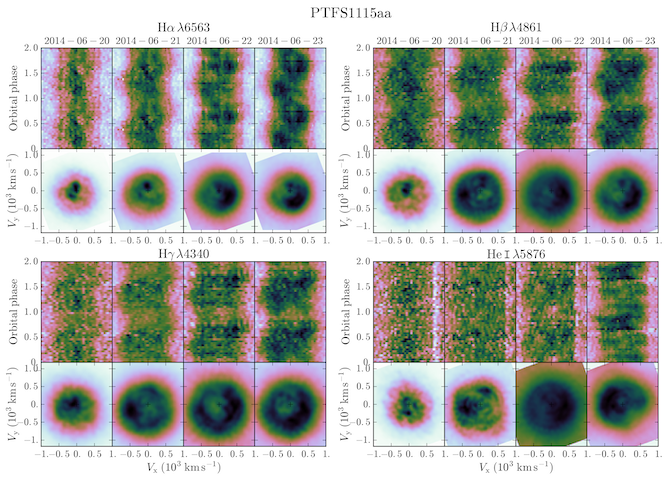}
\end{figure}
\end{landscape}

\begin{landscape}
\begin{figure}
  \caption{Doppler tomograms and trailed spectra for PTFS\,1117an}
  \label{fig:tom_ptfs1117an}
  \centering
    \includegraphics[width=1.35\textheight]{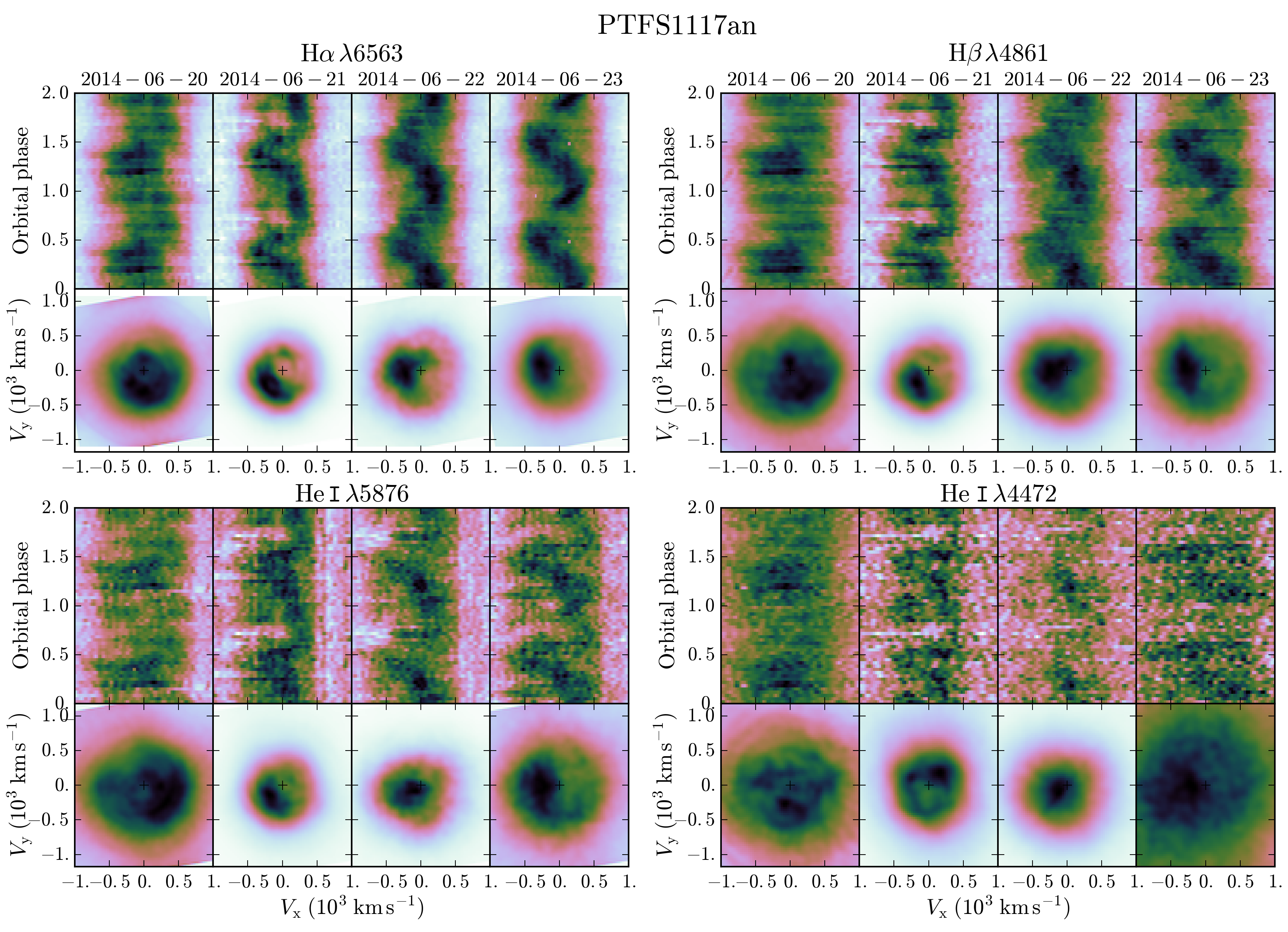}
\end{figure}
\end{landscape}

\begin{landscape}
\begin{figure}
  \caption{Doppler tomograms and trailed spectra for PTFS\,1101af}
  \label{fig:tom_ptfs1101af}
  \centering
    \includegraphics[width=1.35\textheight]{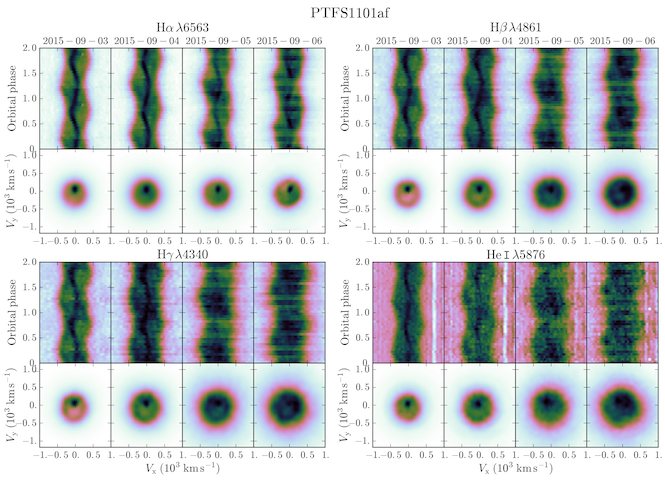}
\end{figure}
\end{landscape}

\begin{landscape}
\begin{figure}
  \caption{Doppler tomograms and trailed spectra for PTFS\,1121b and RZ\,Sge}
  \label{fig:tom_ptfs1121b-rzsge}
  \centering
    \includegraphics[width=1.35\textheight]{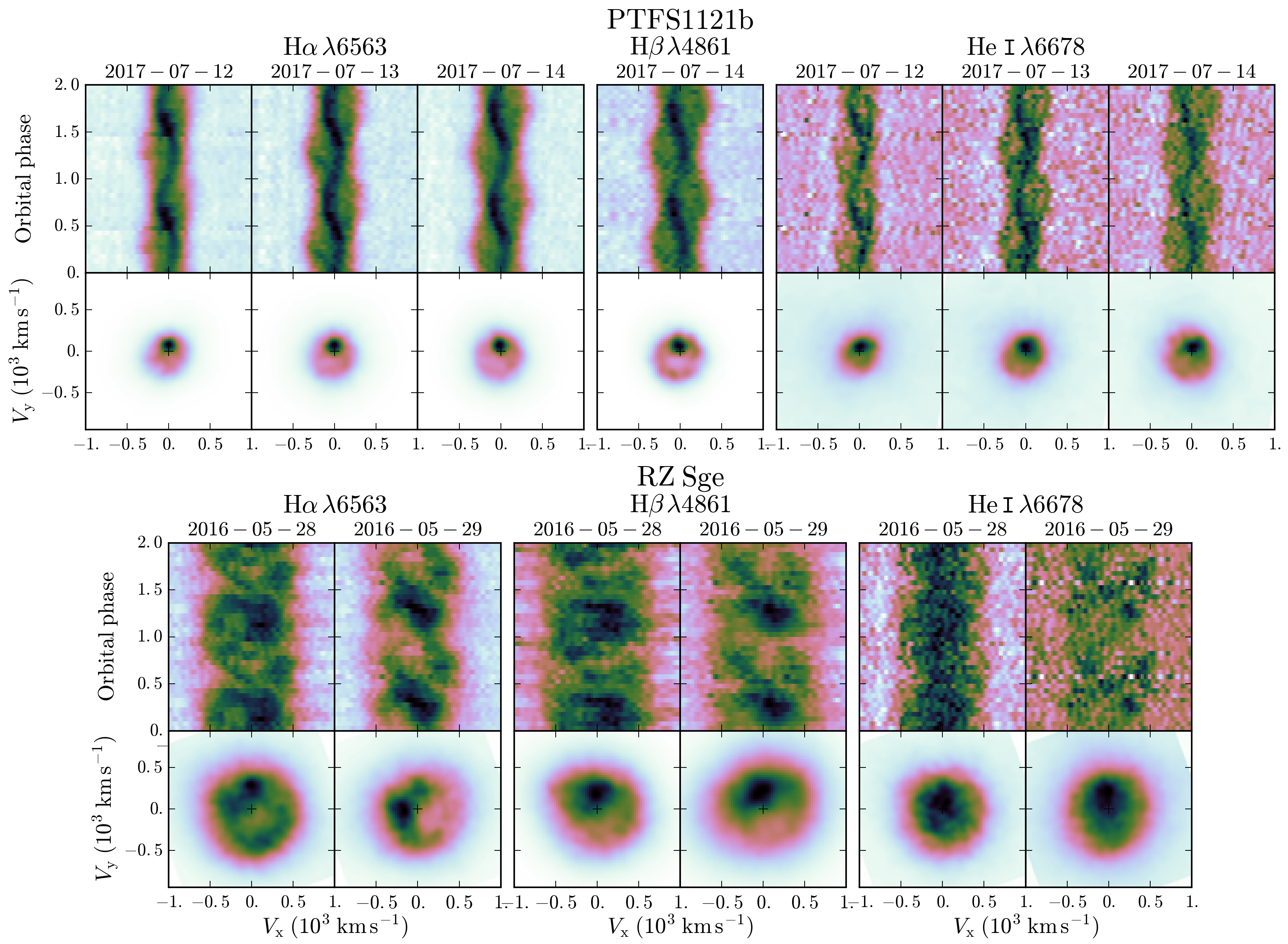}
\end{figure}
\end{landscape}

\begin{figure*}
  \caption{Doppler tomograms and trailed spectra for V795\,Cyg}
  \label{fig:tom_v795cyg}
  \centering
    \includegraphics[width=0.89\textwidth]{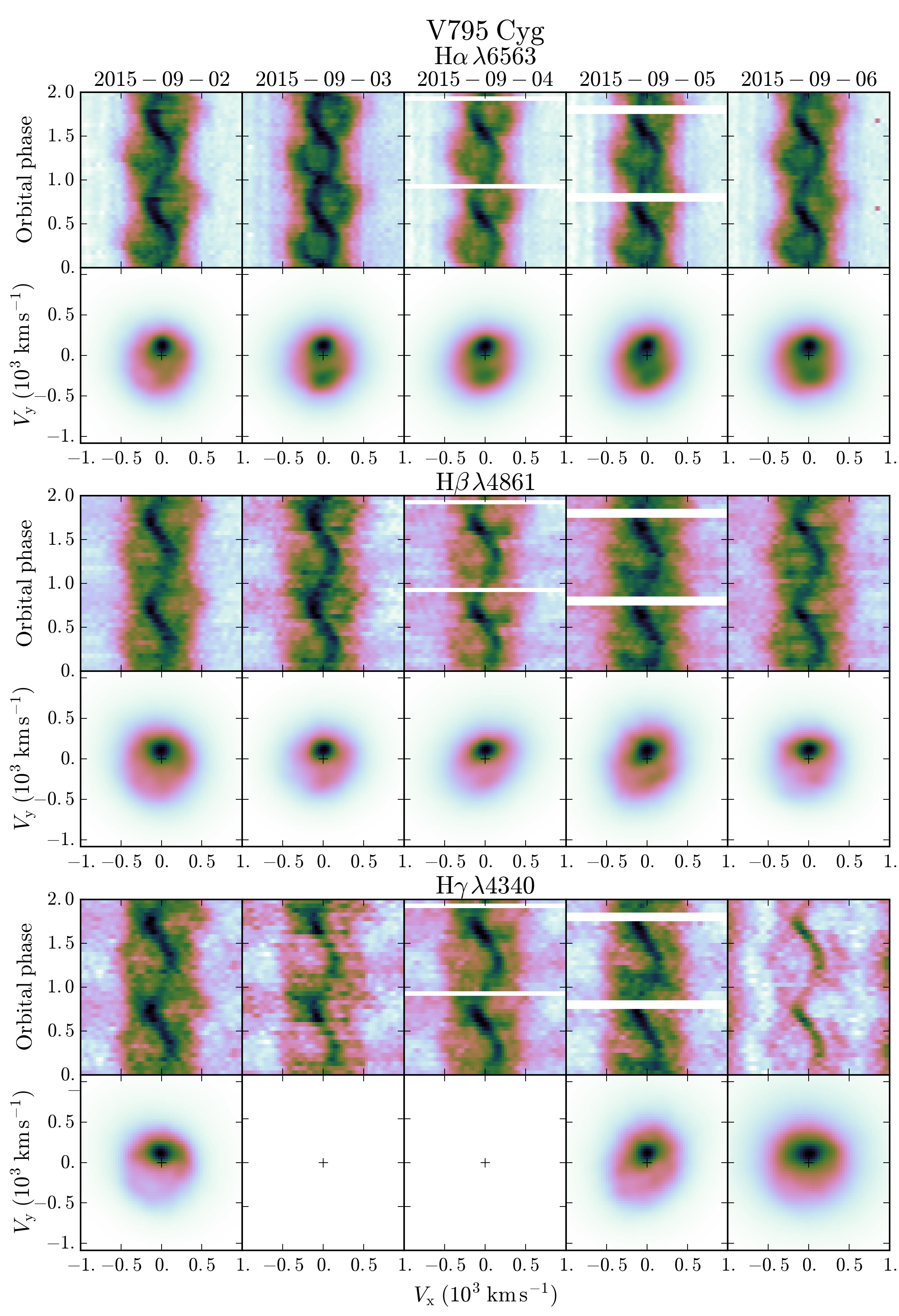}
\end{figure*}

\begin{figure*}
  \caption{Doppler tomograms and trailed spectra for V795\,Cyg (cont.) and AH\,Her}
  \label{fig:tom_v795cyg-ahher}
  \centering
    \includegraphics[width=0.89\textwidth]{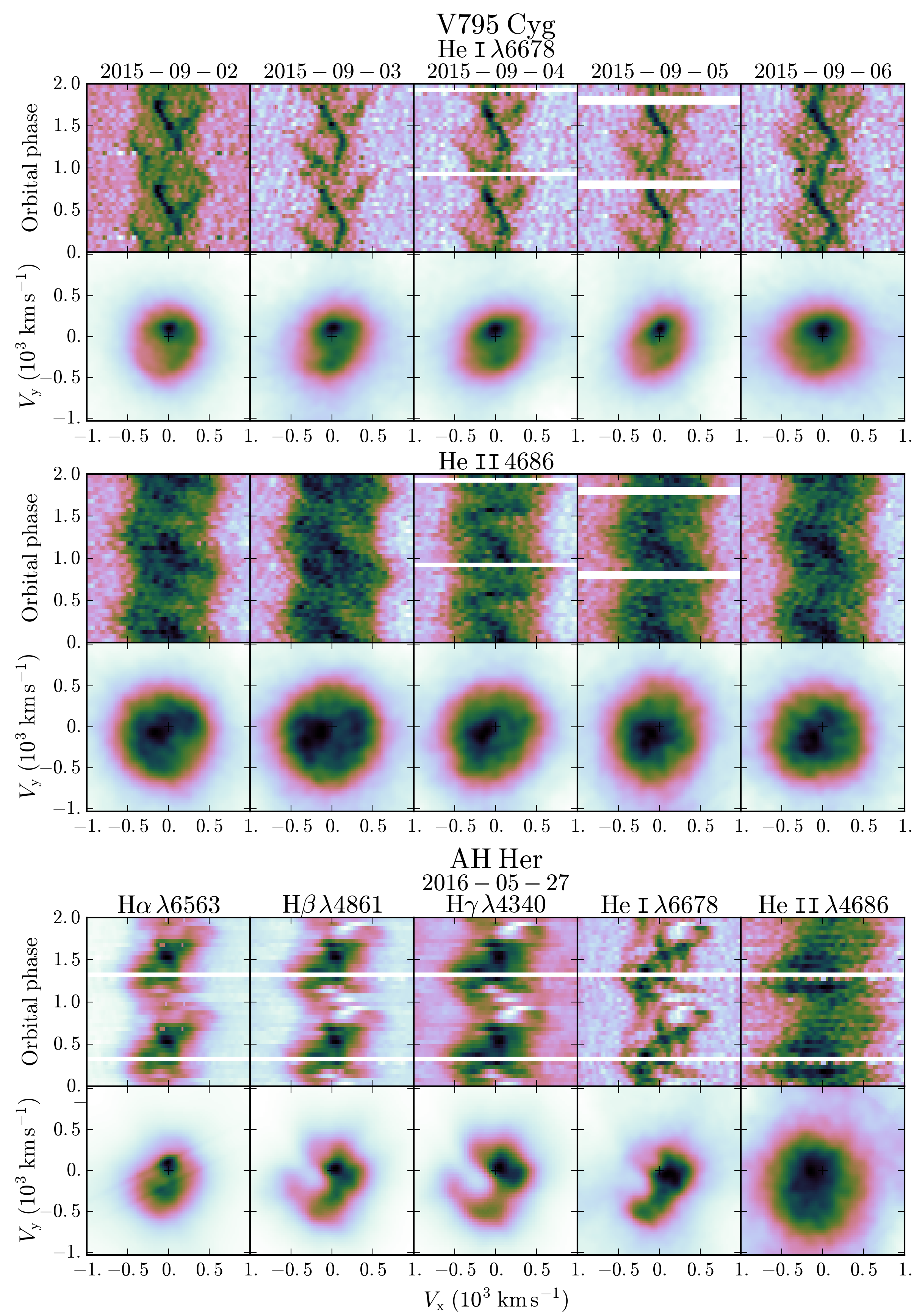}
\end{figure*}

The systems with asymmetric disc structures are:

\begin{itemize}
    \item PTFS\,1115aa (Figure~\ref{fig:tom_ptfs1115aa}); some emission at the middle left of the tomograms consistently shows in the first two nights. A strong emission feature towards the lower right part of the tomograms shows up in the last two nights. The latter component appears stronger and more extended.
        
    \item PTFS\,1117an (Figure~\ref{fig:tom_ptfs1117an}); strong S-waves are present in the trailed spectra that are traceable to an emission region that shifts counterclockwise from the lower right to the top left parts of the tomograms with the progress of the outburst. This behaviour is consistent for H$\alpha$, H$\beta$ and He {\sc i} $\lambda$5876. 
        
    \item PTFS\,1101af (Figure~\ref{fig:tom_ptfs1101af}); very clear S-waves are present in the trailed spectra. The accretion disc and emission from the secondary star are distinguishable in different lines. There is enhanced emission in the trailed spectra at phases 0.0-0.25 towards negative velocities, most notably in the first nights, but this does not result in a localized emission region in the tomograms. 
        
    \item PTFS\,1121b (Figure~\ref{fig:tom_ptfs1121b-rzsge}); the accretion disc and presumed emission from the secondary star is present. There are signs of a third component in the trailed spectra, similarly to PTFS\,1101af, but this is difficult to distinguish in the tomograms. 
        
    \item RZ\,Sge (Figure~\ref{fig:tom_ptfs1121b-rzsge}); a clear disc structure is seen in H$\alpha$ only. Interestingly, the presumed emission from the secondary star appears rotated about 90 degrees counterclockwise in the H$\alpha$ tomogram of the second night (May 29), while remaining at the expected position in other lines. If this is the same emission component it is evidently not from the secondary star. The tomogram of H$\alpha$ on May 28 shows some extended structures toward the top left and bottom right of the tomogram. 
    
    \item V795\,Cyg (Figures~\ref{fig:tom_v795cyg}-\ref{fig:tom_v795cyg-ahher}); the clearest accretion disc structure is present in the H$\gamma$ tomogram of September 2. The accretion disc appears nearly axisymmetric in the first night, but it is manifestly asymmetric in subsequent nights. The trailed spectra display two distinct S-waves that trace back to the secondary star and an emission region towards the lower part of the tomograms. This emission component is clearly visible in the Balmer lines and He {\sc i} $\lambda$6678, and remains roughly in the same position from night to night.
        
    \item AH\,Her (Figure~\ref{fig:tom_v795cyg-ahher}); similar emission components as in V795\,Cyg are seen in the trailed spectra and tomograms, in particular the secondary star and an enhanced emission towards the lower left of the tomograms. 
        
\end{itemize}

\subsection{Tomograms with spiral structure}
\label{ssec:res-sdw} 

In this Section, we report on the systems that show spiral structure in their tomograms. We describe the significance maps we obtained via bootstrapping of selected tomograms (Figure~\ref{fig:bootstraps}). Additional tomograms can be found in Appendix~\ref{sec:dopplermaps}. The systems with evidence of spiral structure are:
\begin{itemize}
        
    \item PTFS\,1200o; exhibits evidence of strong spiral strucure in its H$\beta$ and H$\alpha$ tomograms and especially in the He {\sc ii} $\lambda$4686 line. In the Balmer lines, the secondary star seems visible in all tomograms while the analogous component in He {\sc ii} $\lambda$4686 appears shifted towards the centre of the tomogram.
    
\end{itemize}



\begin{figure*}
  \caption{Trailed spectra, Doppler tomograms and significance maps for selected emission lines and systems.}
  \label{fig:bootstraps}
  \centering
    \includegraphics[width=0.9\textwidth]{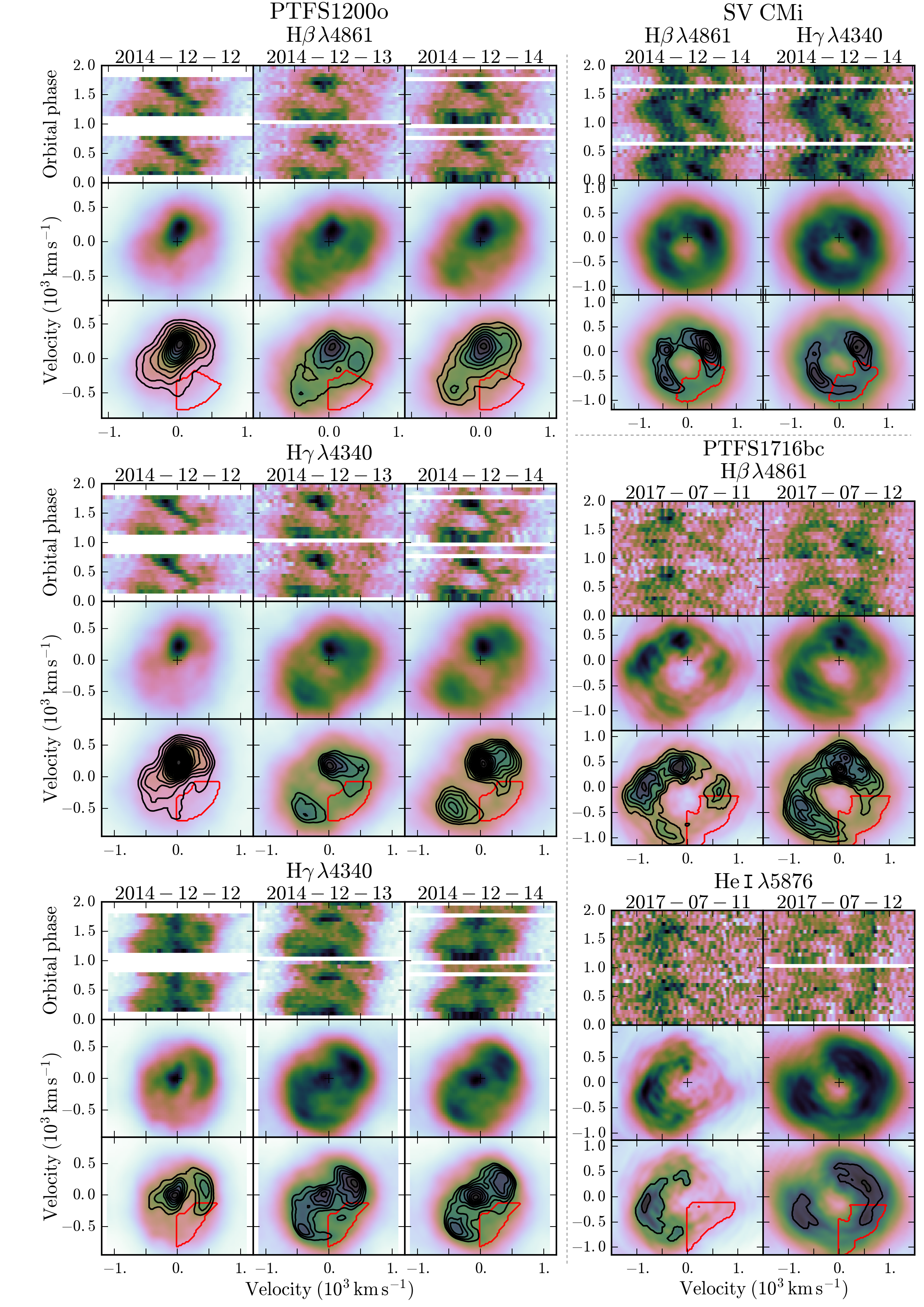}
\end{figure*}

        
    
    
    The significance maps reveal asymmetric discs in the tomograms, with strong emission from the secondary star. In H$\beta$, the significance of the spiral arms cannot be confirmed. In H$\gamma$, the arm in the lower left is established in the significance maps on December 13 and, more strongly, on December 14. The arm at the upper right is confused with the emission from the secondary star. In He {\sc ii} $\lambda$4686, the significance maps confirm the robust detection of spiral density waves, especially in the later two nights. The arm at the lower left grows stronger with the progress of the outburst, while the arm at the upper right appears less extended every night with maximum strength on December 13. 

\begin{itemize}
        
    \item SV\,CMi; shows spiral structure on December 14 in H$\beta$ and H$\gamma$, with some evidence of a bright spot. Tomograms on December 13 show a clear two-armed structure, but data at many phases are missing and we will not further discuss those tomograms. The {\sc i} lines are too faint to compute Doppler maps.
    
    The significance maps support the solid detection of spiral waves in H$\beta$ and H$\gamma$. The lower left arm is more extended in H$\gamma$, and part of it was included in the accretion disc region, resulting in slightly lower values of the significance parameter. There is also evidence of a bright spot.
    
    \item PTFS\,1716bc; this is a very faint system. There is evidence of spiral structure in the tomograms, especially on July 12. The spiral structure is best visible in H$\beta$ and He {\sc i} $\lambda$5876, while the tomograms in H$\alpha$ and H$\gamma$ show fainter asymmetric discs.
    
    The significance maps in H$\beta$ are patchy and irregular, although reaching values of $m=6$. In He {\sc i}, the significance parameter is low, with no regions characterised by $m \ge 3$. We conclude that the detection of spiral waves is not yet robust in this system and will require higher signal-to-noise coverage during a future outburst. 
        
\end{itemize} 


\section{Discussion}
\label{sec:discussion}

The premise of our research was to determine the prevalence of spiral density waves in accretion discs in outburst, and thereby establish their role to the physics of hot discs, and in particular to their role in the transport of angular momentum through hot viscous discs. 

It is worth noting that the spiral structures detected in Doppler tomograms are obtained from enhanced line emission, and are not necessarily shocks. Emission lines typically form in superficial layers of the disc, and assessing whether or not there is underlying vertical structure through the bulk of the disc is a difficult task. Along these lines, there are critical views about the interpretation of tomograms such as \citet{2001AcA....51..295S}, and \citet{2002MNRAS.330..937O}, who propose that the emission in the tomograms originates in irradiated tidally-disturbed regions in the disc, not in shocks. As noted by \citet{2001AcA....51..295S}, it is often assumed in numerical simulations that disc surface density and line emissivity are the same thing which may not be true, as very few radiative transfer studies have been made. Future simulations should incorporate a proper treatment of radiative processes in order to elucidate if spiral shocks are detectable with Doppler tomography. 

\subsection{Expectations on spiral density waves in outburst}
\label{ssec:dis-outburst}

Our study reveals a low turnout of spiral density waves in CVs during dwarf nova outburst. Although these results are not entirely unprecedented, e.g. OY\,Car \citep{1996A&A...308...97H}, the detection of spiral density waves was expected based on a number of arguments. 

First, the technique of Doppler tomography is well suited to map spiral density waves down to intensity contrast levels of about 8\% with respect to the disc, opening angles greater than $\phi < -73^{\circ}$ (for spectral and temporal resolution comparable to our data) and signal-to-noise ratios (SNR) as low as 5 -- 10 in the individual spectra \citep{sdw_detectability}. Spiral shocks have been detected to contribute 15\% to the emission of the tomogram (e.g. \citealt{2001ApJ...551L..89G}) and to have large opening angles during outburst ($\phi \approx -55^{\circ}$ in IP\,Peg, \citealt{1997MNRAS.290L..28S}; \citealt{1999MNRAS.307...99S}). The quality of our data is well above the SNR threshold (SNR = 15 -- 60); note that we found spiral structure in the faintest system, i.e. PTFS\,1716bc. 

Second, the detection of spiral density waves is repeatable. Tomograms of IP\,Peg showed spiral density waves in at least five distinct outbursts (\citealt{1990ApJ...349..593M}, \citealt{1996MNRAS.281..626S}, \citealt{1997MNRAS.290L..28S}, \citealt{1999MNRAS.306..348H}, \citealt{2000MNRAS.313..454M}). V347\,Pup and EC21178--54 also showed spiral waves at different epochs (\citealt{1998MNRAS.299..545S}, \citealt{2005MNRAS.357..881T}; and \citealt{Khangale:2013uq}, \citealt{RuizCarmona_ec} respectively). This repeatability suggests that, in a given system, spiral structure is always detectable in accretion discs in outburst, especially considering the historic scarcity of phase-resolved spectroscopic data during dwarf nova outbursts. Our observational strategy has now solved the difficulties of finding systems in outburst, removing the obstacles for collecting this type of data. 

Third, the mechanisms that trigger the formation of spiral density waves prevail in CVs. Tidal interaction is a consequence of the non-axisymmetric gravitational potential characteristic of compact binaries, and is especially at work in outbursting discs as they are hot and large in size and can therefore more easily reach a resonance radius (\citealt{1988MNRAS.232...35W}, \citealt{1994MNRAS.268...13S}). A substantial amount of numerical simulations show the development of spiral density waves (e.g. \citealt{1986MNRAS.221..679S}, \citealt{1998MNRAS.297L..81A}, \citealt{1999MNRAS.307...99S}, \citealt{2005ApJ...632..499L}, \citealt{2010MNRAS.405.1397M}, \citealt{2017ApJ...841...29J}). 

As a counterargument, even though spiral waves may develop in outbursting discs, the spiral structure is not necessarily observable, either due to low contrast with the disc, or because the emission regions may be obscured by self-absorption. 


\subsection{Overview of system parameters}
\label{ssec:dis-pars}

In this Section, we examine the system parameters of compact binaries to understand if any of them favour the detection of spiral density waves. For this purpose, we have gathered and calculated the information presented in Table~\ref{tab:ours_pars}, using the methods and expressions in Appendix~\ref{sec:calc_pars}. We also present the physical parameters of systems for which there are published claims of spiral structure detection in their tomograms (Table~\ref{tab:pub_pars}). For consistency, we have excluded studies that use techniques other than Doppler tomography, e.g. IY\,UMa with light curve fitting \citep{2008ARep...52..815K} or V348\,Pup with eclipse mapping \citep{2016MNRAS.457..198S}. Similar to the treatment of our results, we distinguish systems where spiral shocks are at the expected phases (i.e. similar to those seen in IP\,Peg and U\,Gem) from systems that display enhanced emission from asymmetries in their disc. We briefly comment on the tomograms from literature in Appendix~\ref{sec:comm_sdws}. We represent some of these parameters against orbital period in Figure~\ref{fig:orbital_pars}, and discuss their potential influence in the detection of spiral waves.

\subsubsection{Inclination}
\label{sssec:i}

In highly inclined systems, the disc is seen edge-on and the continuum from the optically thick disc can be hidden from view. At lower inclinations many discs in outburst display absorption lines \citep{1990ApJ...349..593M}. In highly inclined systems, emission lines formed above the disc in a chromosphere-type configuration may become more visible due to the foreshortening of the disc continuum. Many of the systems with published tomograms that exhibit spiral density waves are eclipsing and therefore at high inclination, and show emission lines during the outburst phases. The two eclipsing systems in our sample, PTFS\,1200o and PTFS\,1716bc, show evidence of spiral structure in their tomograms. Hence, the systems' inclination is of importance in the detection of spiral shocks, although not a prerequisite (Figure~\ref{fig:orbital_pars}a). SS\,Cyg \citep{1996MNRAS.281..626S} and SV\,CMi show evidence of spiral structure even though they are not eclipsing ($i \approx 55^{\circ}$), but their outburst spectra do display pure emission lines. 

On the other hand, the eclipsing systems V2051\,Oph, V2779\,Oph, HT\,Cas and CRTS\,J0359 do not display a clear spiral structure, but instead present discs with asymmetric emission at phases other than those expected for spiral density waves. For instance, CTRS\,J0359 presents absorption components in the centre of the lines \citep{2018AJ....155..232L}. V2779\,Oph seems to have absorption components that vary in strength with orbital phase \citep{2010AIPC.1273..346Y}; but this is not the case for HT\,Cas nor CRTS\,J0359. HT\,Cas shows emission components that visible at certain phases in the trailed spectra \citep{2016A&A...586A..10N}, but do not display clear S-waves, perhaps due to self-obscuration. 

For highly-inclined systems, the line-of-sight can intersect regions of variable optical depth if the scale height at the outer rim of the disc is phase dependent. This can effectively hide emission from parts of the disc, resulting in asymmetries in the tomograms. If scale height scales with e.g. temperature, it is plausible that discs exhibit variable scale heights, e.g. around the hot spot, leading to self-obscuration.

\subsubsection{Mass ratio}
\label{sssec:q}

Accretion discs in compact binaries are susceptible to Lindblad resonance in cases where the mass ratio allows for the discs to grow up to a resonance radius: (\citealt{1979MNRAS.186..799L} -- resonance 2:1 if  $q \lsim 0.1$; \citealt {1988MNRAS.232...35W} -- resonance 3:1 if $q \lsim 0.25$). Spiral density waves can be excited as a result of this type of tidal interaction \citep{1979MNRAS.186..799L}. At larger mass ratios, the Lindblad 3:1 resonance can be important if the disc is hot \citep{1994MNRAS.268...13S}.

Figure~\ref{fig:orbital_pars}b shows how CRTS\,J1111, WZ\,Sge and PTFS\,1716bc, which show spiral structure in their tomograms, are in the mass ratio range where tidal resonances can occur in the disc. In contrast, PTFS\,1115aa and, especially, CRTS\,J039 are characterised by mass ratios too large to be susceptible to tidal resonance, although they both exhibit superhumps. There is no correlation between mass ratio and the occurrence of detectable spiral structures for systems with orbital periods longer than 3 hours.

\subsubsection{Mass of the primary}
\label{sssec:m1}

Spiral density structures are more prominent in Doppler tomograms of He {\sc ii} $\lambda$4686 (e.g. \citealt{2002MNRAS.332..814M}). This could point to a higher-mass of the primary white dwarf, since accretion into a deeper potential can produce more ionizing radiation able to enhance He {\sc ii} emission. The same effect can be obtained by higher mass-accretion rates or by weakly magnetized flows causing accretion shocks. 

In Figure~\ref{fig:orbital_pars}c, we can see how CRTS\,J1111, V406\,Vir, IP\,Peg, U\,Gem and PTFS\,1200o harbour more massive white dwarfs compared to other systems at their orbital periods. The systems that exhibit spiral structure in our sample, i.e. PTFS\,1200o, SV\,CMi and PTFS\,1716bc, present He {\sc ii} emission lines in their spectra. It should be noted that masses of the stellar components are hard to estimate, and assuming an average value  $\langle M_1 \rangle = 0.75$ is common practice \citep{2005PASP..117.1204P}. 

\subsubsection{Mass transfer rate}
\label{sssec:mdot}

\begin{figure*}
  \caption{System parameters as a function of orbital period. Circles show systems studies here, squares indicate systems reported in the literature. Fill styles indicated tomographic difference ranging from detected spiral (fully filled symbols), to non-structured asymmetries (half-filled symbols) to no structure (open symbols). 
 \textit{(a)}: Inclination; the dashed horizontal line separates eclipsing from non-eclipsing systems. \textit{(b)}: Mass ratio; dashed horizontal lines indicate upper limits for tidal resonances as indicated. \textit{(c)}: Mass of the white dwarf. \textit{(d)}: Mass accretion rate estimated using mass transfer rates from the secondary and duty cycle. The shown relation has been derived for systems below the gap, but extrapolated for longer period systems. \textit{(e)}: Mass accretion rate from outburst luminosity. Ranges of values of the mass accretion rate from the literature are also represented when available in panels \textit{(d)} and \textit{(e)} by dashed, vertical lines.}
  \label{fig:orbital_pars}
  \centering
    \includegraphics[width=0.85\textwidth]{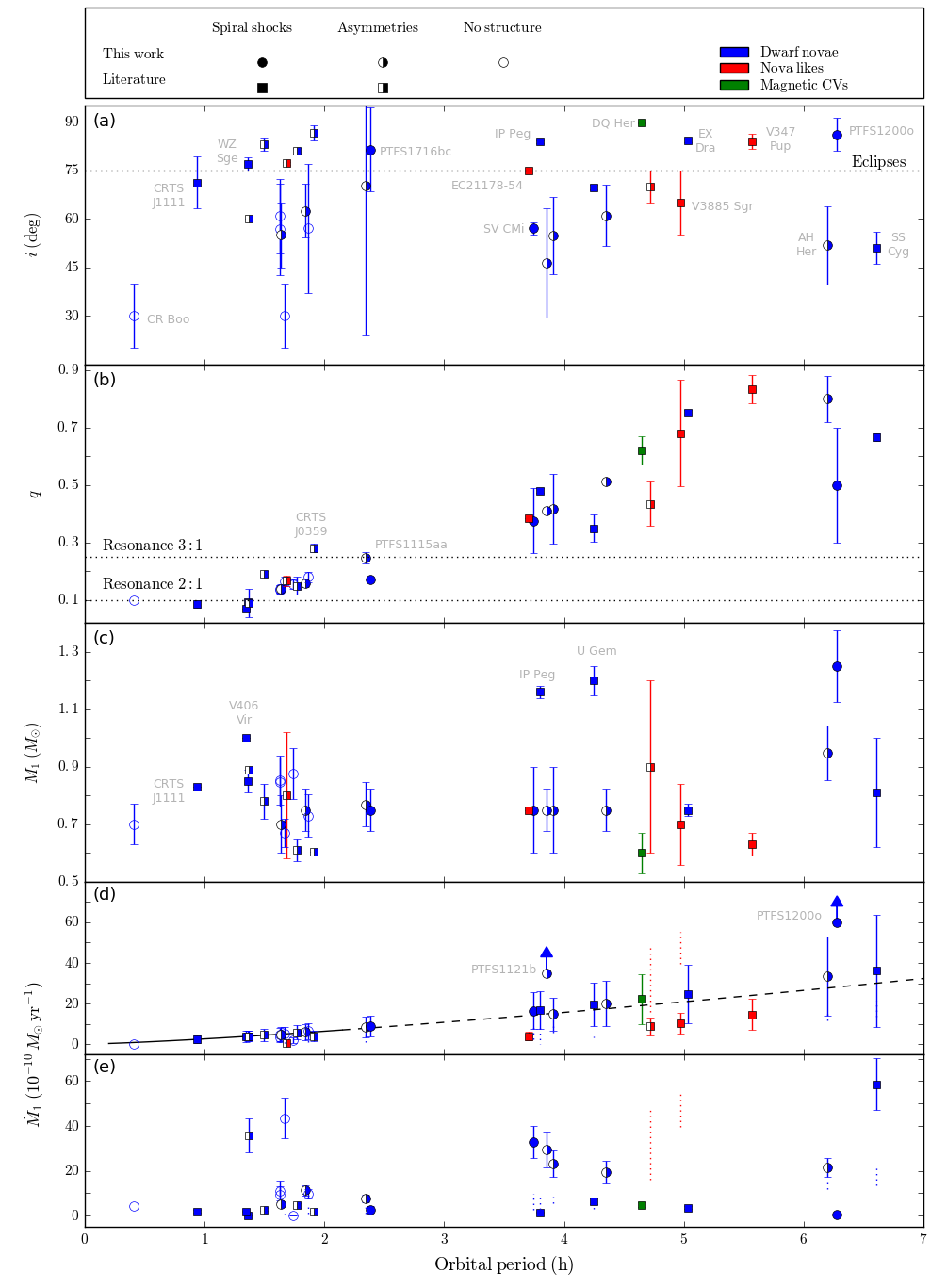}
\end{figure*}

Finally, we study how the detection of spiral waves relates to the mass accretion rate. We estimated the mass accretion rate in two ways (see Appendix~\ref{ssec:acc_rate} for details on the calculations): using the mass transfer rate from the secondary and duty cycles in Figure~\ref{fig:orbital_pars}d, and using the outburst luminosity in Figure~\ref{fig:orbital_pars}e. 

In Figure~\ref{fig:orbital_pars}d, data points generally follow the equations in Appendix~\ref{ssec:acc_rate}. Novalikes, for which we assumed to be always in the high state and equated their mass accretion rate to their mass transfer, fall below the general trend. PTFS\,1121b and PTFS\,1200o, for which realistic duty cycles have been reported \citep{2016MNRAS.456.4441C}, are outliers on the high side. Additional studies of duty cycles based on short-cadence, dense coverage light curves are necessary to draw conclusions. Current projects such as the Zwicky Transient Facility \citep{2019PASP..131a8002B} will be ideal for this. 

In Figure~\ref{fig:orbital_pars}e, we note that many systems with evidence of spiral structure are characterised by mass accretion rates lower than $10^{-9} M_{\rm{\odot}}\, \rm{yr}^{-1}$, although the systematic uncertainties in these calculations are large. If the spiral structure detected in the tomogram were related to spiral shocks that lead to enhanced angular momentum transport one would expect these systems to show {\sl high} accretion rates during outburst.



\begin{landscape}
\begin{table}
  \centering
  \caption{System parameters of the observed systems in this study. Columns are orbital period, superhump period, mass ratio ($q$), mass of the secondary star, mass of the primary star, inclination, distance, duty cycle {$dc$}, mass transfer rate from the secondary star, mass accretion rate onto the primary estimated from the mass transfer rate, from the outburst luminosity and from the literature. Details of the calculations are given in Section~\ref{sec:calc_pars}.}
  \label{tab:ours_pars}
  \renewcommand*{\arraystretch}{1.3}
  
  \begin{threeparttable}[b]
  \begin{tabular}{l S[table-format=1.3]@{\hskip 0.3in}  S[table-format=1.3]@{\hskip 0.3in} 
                    c@{\hskip 0.1in}c@{\hskip 0.2in}c@{\hskip 0.2in}
                    c@{\hskip 0.3in}
                    c@{\hskip 0.3in}c@{\hskip 0.3in} 
                    c@{\hskip 0.2in}c@{\hskip 0.2in}
                    c@{\hskip 0.2in}c}\toprule
 	
    \hline \hline 
    \multirow{2}{1.2cm}{\centering{System}} 
    & $P_{\rm{orb}}$   & $P_{\rm{SH}}$  
    & \multirow{2}{1.0cm}{\centering{$q$}}     
    & $M_2$  &  $M_1$  &  $i$  &  $d$\tnote{h} 
    & \multirow{2}{0.8cm}{\centering{$dc$}}
    & {$\dot{M}_2$}\tnote{j} & {$\dot{M_1}_{\rm{,}\,dc}$}\tnote{k} & 
    {$\dot{M_1}_{\rm{,}\,out}$}\tnote{l} & {$\dot{M_1}_{\rm{,}\,ref}$}\tnote{m} \\[3pt]
    
    \cline{10-13}
    
    & [h] & [h] &  & [$M_{\rm{\odot}}$] & [$M_{\rm{\odot}}$] & [deg] & [pc] &
    & \multicolumn{4}{c}{[$10^{-10} M_{\rm{\odot}}\, \rm{yr}^{-1}$]} \\[3pt]
    
    \hline \midrule
    
    PTFS\,1118m  & 1.628\tnote{a}  & 1.678\tnote{E} 
               & 0.13\tnote{c}   & 0.12\tnote{e}    & 0.85 
               & 57 $\pm$ 14\tnote{g}   & $1104^{+192}_{-142}$ 
               & 0.06\tnote{i}          & 0.3 $\pm$ 0.2 
               & 4.8 $\pm$ 3.3   & 11.1 $\pm$ 4.5  \\
                        
    PTFS\,1122s  & 1.625\tnote{b}  & 1.666\tnote{F} 
               & 0.14\tnote{c}   & 0.12\tnote{e}    & 0.85 
               & 61 $\pm$ 12\tnote{g}   & $396^{+62}_{-47}$ 
               & 0.09\tnote{P}          &  0.3 $\pm$ 0.2 
               & 3.2 $\pm$ 2.2   & 9.4 $\pm$ 3.5 \\
                        
 	PTFS\,1115aa & 2.341\tnote{A}  & 2.524\tnote{G}
 	           & 0.25\tnote{K}   & 0.19\tnote{K}    & 0.81\tnote{K} 
 	           & 70 $\pm$ 47\tnote{g}   & $772^{+31}_{-29}$ 
 	           & 0.11\tnote{i}          & 0.9 $\pm$ 0.5
 	           & 8.4 $\pm$ 5.0   & 7.6 $\pm$ 1.6    & 1.0 -- 4.3\tnote{K} \\
 	
 	PTFS\,1117an & 1.837\tnote{B}  & 1.906\tnote{H} 
 	           & 0.16\tnote{B}   & 0.12\tnote{e}    & 0.75 
 	           & 63 $\pm$ 08\tnote{g}    & $536^{+11}_{-10}$ 
 	           & 0.07\tnote{P}           & 0.4 $\pm$ 0.2 
 	           & 6.0 $\pm$ 3.9   & 11.3 $\pm$ 2.3  \\
 	
 	PTFS\,1200o  & 6.279\tnote{a}  & 
 	           & 0.50\tnote{d}   & 0.62 & 1.24\tnote{d} 
 	           & 86 $\pm$ 05\tnote{d}    & $1212^{+482}_{-268}$ 
 	           & 0.04\tnote{P}   & 21.6 $\pm$ 11.4  
 	           & 540 $\pm$ 316   & 0.6  $\pm$ 0.5\\
 	
 	SV\,CMi     & 3.744\tnote{C}  & 
 	           & 0.38\tnote{K}   & 0.28\tnote{K}  & 0.75\tnote{K} 
 	           & 57 $\pm$ 02\tnote{K}  & $455^{+20}_{-18}$\tnote{ K} 
 	           & 0.25\tnote{i}   & 4.3 $\pm$ 2.0
 	           & 16.5 $\pm$ 9.1  & 32.8 $\pm$ 7.2 & 2.4 -- 9.5\tnote{K} \\
 	
 	PTFS\,1119h  & 1.668\tnote{C}  & 1.733\tnote{C} 
 	           & 0.16\tnote{K}   & 0.11\tnote{K}    & 0.67\tnote{K} 
 	           & 30 $\pm$ 10\tnote{K}   & $527^{+13}_{-12}$ 
 	           & 0.06\tnote{i}   & 0.3 $\pm$ 0.2 
 	           & 5.2 $\pm$ 3.5   & 43.5 $\pm$ 9.0 & 0.6 -- 0.8\tnote{K}\\
 	
 	PTFS\,1101af & 3.912\tnote{C}  & 
 	           & 0.42\tnote{K}   & 0.31\tnote{K} & 0.75\tnote{K} 
 	           & 55 $\pm$ 12\tnote{g}   & $438^{+35}_{-30}$ 
 	           & 0.32\tnote{P}   & 4.8 $\pm$ 2.3 
 	           & 14.8 $\pm$ 8.2  & 23.1 $\pm$ 5.9 & 5.6 --10.3\tnote{K}\\
 	
 	V795\,Cyg   & 4.351\tnote{C}  & 
 	           & 0.51            & 0.39\tnote{e} & 0.75\tnote{f} 
 	           & 61 $\pm$ 09\tnote{g}    & $655^{+52}_{-45}$ 
 	           & 0.33\tnote{i}   & 6.7 $\pm$ 3.3
 	           & 20.3 $\pm$ 11.2 & 19.3 $\pm$ 4.9 \\
 	
 	AH\,Her     & 6.195\tnote{C} &  
 	           & 0.80\tnote{C}   & 0.76\tnote{C} & 0.95\tnote{C} 
 	           & 52 $\pm$ 03\tnote{g}   & $324^{+3}_{-3}$ 
 	           & 0.62\tnote{i}   & 20.7 $\pm$ 10.9 
 	           & 33.4 $\pm$ 19.4 & 21.4 $\pm$ 4.3  & 11.9 -- 15.1\tnote{K}\\
 	
 	PTFS\,1113bc & 0.409\tnote{D}  & 0.424\tnote{I}
 	           & 0.10\tnote{I}   & 0.07\tnote{N} & 0.65 
 	           & 30 $\pm$ 10\tnote{N}   & $337^{+44}_{-35}$\tnote{ N} 
 	           & 0.74\tnote{Q}      & 0.003 $\pm$ 0.004 
 	           & 0.005 $\pm$ 0.006  & 4.4 $\pm$ 1.4 & 0.04 -- 0.12\tnote{N}\\
 	
 	PTFS\,1215t  & 1.735\tnote{b}  & 1.795\tnote{J} 
 	           & 0.16\tnote{c}   &  0.14\tnote{e}  & 0.87 
 	           &  &  
 	           & 0.18\tnote{P}   & 0.4 $\pm$ 0.2 
 	           & 2.0 $\pm$ 1.3   &  \\
 	
 	RZ\,Sge     & 1.639\tnote{C}  & 1.694\tnote{C} 
 	           & 0.14\tnote{C}   & 0.10  & 0.70\tnote{L} 
 	           & 55 $\pm$ 10\tnote{B}    & $256^{+9}_{-8}$ 
 	           & 0.06\tnote{i}           & 0.3 $\pm$ 0.2
 	           & 4.9 $\pm$ 3.3   & 5.1 $\pm$ 1.1  & \\
 	
 	V503\,Cyg   & 1.866\tnote{C}  & 1.944\tnote{C} 
 	           & 0.18\tnote{K}   & 0.13\tnote{K}   & 0.73\tnote{K} 
 	           & 57 $\pm$ 20\tnote{B}    & $440^{+17}_{-16}$ 
 	           & 0.07\tnote{i}           & 0.4 $\pm$ 0.3
 	           & 6.4 $\pm$ 4.1   & 9.6 $\pm$ 2.1 & 1.0 -- 4.4\tnote{K} \\
 	
 	PTFS\,1121b  & 3.854\tnote{C}  & 
 	           & 0.41  & 0.31\tnote{e} & 0.75\tnote{f} 
 	           & 46 $\pm$ 17\tnote{B}  & $585^{+53}_{-45}$ 
 	           & 0.07\tnote{P}         & 4.5 $\pm$ 2.2
 	           & 64.7 $\pm$ 35.6  & 29.5 $\pm$ 8.0 \\
 	
 	PTFS\,1716bc & 2.387\tnote{a}  & 
 	           & 0.17  & 0.13\tnote{e} & 0.75\tnote{f}
 	           & 82 $\pm$ 12\tnote{g}  & $1828^{+674}_{-388}$ 
 	           & 0.11\tnote{i}         & 1.0 $\pm$ 0.5
 	           & 8.9 $\pm$ 5.3   & 2.5 $\pm$ 1.9\\[3pt]
 	
 	\hline \bottomrule
  \end{tabular}
  
  \begin{tablenotes}[para, flushleft]
  \footnotesize
  {\bf Notes:}\hspace{-0.4cm}
     \item[a] Orbital period obtained in this work via light curve fitting (see Section~\ref{ssec:lcfitting}). \hspace{-0.5cm}
     \item[b] Orbital period obtained from superhump period \citep{2009MNRAS.397.2170G}. \hspace{-0.5cm}
     \item[c] Mass ratio estimated from the period excess \citep{2005PASP..117.1204P}. \hspace{-0.5cm}
     \item{d} Estimated from light curve fitting with \textsc{lcurve}. \hspace{-0.5cm} \item[e] Mass of the secondary star estimated from the broken-power-law semi-empirical donor sequence \citep{2011ApJS..194...28K} and the requirement that it fills its Roche lobe \citep{1972ApJ...175L..79F}.\hspace{-0.5cm}
     \item[f] Assumed mean value $\langle M_1 \rangle = 0.75$ \cite{2005PASP..117.1204P}. \hspace{-0.5cm}
     \item[g] With the absolute magnitude of outburst estimated from the orbital period \citep{2011MNRAS.411.2695P}, the apparent magnitude and the distance, we derive the inclination from the correction of the absolute magnitude for a flat, limb-darkened accretion disc \citep{1980AcA....30..127P}. \hspace{-0.5cm}
     \item[h] Distances from {\it Gaia} parallaxes (\citealt{2016A&A...595A...1G}, \citealt{2018A&A...616A...1G}, \citealt{2018A&A...616A...9L}). \hspace{-0.5cm} 
     \item[i] Duty cycle estimated as a function of the orbital period \cite{2016MNRAS.456.4441C}. This expression is only valid for systems below the period gap. \hspace{-0.5cm}
     \item[j] Mass transfer rate from the secondary star estimated as a function of the orbital period \citep{1984ApJS...54..443P}. \hspace{-0.5cm}
     \item[k] Upper limit on the secular mass accretion rate from the mass transfer rate and the duty cycle, neglecting accretion during quiescence. \hspace{-0.5cm}
     \item[l] Limit on the maximum instantaneous mass accretion rate during outburst estimated from the excess luminosity, assuming that half of the energy available is released as radiation. \hspace{-0.5cm}
     \item[m] Range of accretion rates published in the literature.
     
  \hfill\parbox[t]{21.8cm}{
  {\bf References:}
     [A] \citealt{2018Ap.....61...64S}
     [B] \citealt{2011MNRAS.411.2695P}
     [C] \citealt{2003A&A...404..301R}
     [D] \citealt{1997ApJ...480..383P}
     [E] \citealt{vsnet-18921}
     [F] \citealt{cvnet-3358}
     [G] \citealt{2019PASJ...71L...1K}
     [H] \citealt{2009PASJ...61S.395K}
     [I] \citealt{2016PASJ...68...64I}
     [J] \citealt{vsnet-20769}
     [K] \citealt{2018A&A...617A..26D} and references therein
     [L] \citealt{2017MNRAS.466.2855P}
     [M] \citealt{2005PASP..117.1204P}
     [N] \citealt{2007ApJ...666.1174R}
     [P] \citealt{2016MNRAS.456.4441C}
     [Q] \citealt{1987ApJ...313..757W}
     }

    \end{tablenotes}

\end{threeparttable}
\end{table}
\end{landscape}


\begin{landscape}
\begin{table}
  \centering
  \caption{System parameters of the systems with detected spiral structures. Columns are orbital period, mass ratio ($q$), mass of the secondary star, mass of the primary star, inclination, distance, duty cycle ($dc$), mass transfer rate from the secondary star, mass accretion rate onto the primary estimated from the mass transfer rate, from the outburst luminosity and from the literature, state at which spiral waves were found and references. Details of the calculations are given in Section~\ref{sec:calc_pars}.}
  \label{tab:pub_pars}
  \renewcommand*{\arraystretch}{1.3}
  
  \begin{threeparttable}[b]
  \begin{tabular}{lc S[table-format=1.3]  cccccc  cccc  c ll}\toprule
 	
    \hline \hline 
    \multirow{2}{1.0cm}{System} 
    & \multirow{2}{0.7cm}{\centering{Class\tnote{a}}} 
    & $P_{\rm{orb}}$    
    & \multirow{2}{0.3cm}{\centering{$q$}}     
    & $M_2$  &  $M_1$  &  $i$  &  $d$\tnote{b} 
    & \multirow{2}{0.6cm}{\centering{$dc$\tnote{c}}}
    & {$\dot{M}_2$}\tnote{d} & {$\dot{M_1}_{\rm{,}\,dc}$}\tnote{e} & 
    {$\dot{M_1}_{\rm{,}\,out}$}\tnote{f} & {$\dot{M_1}_{\rm{,}\,ref}$}\tnote{f} 
    && \multicolumn{2}{c}{Spiral density waves} \\[3pt]
    
    \cline{5-6}
    \cline{10-13}
    \cline{15-16}
    
    & & [h] &  & \multicolumn{2}{c}{[$M_{\rm{\odot}}$]} & [deg] & [pc] &
    & \multicolumn{4}{c}{[$10^{-10} M_{\rm{\odot}}\, \rm{yr}^{-1}$]} 
    & & State\tnote{g}  & References\\[3pt]
    
    \hline \midrule

    IP\,Peg     & UG      & 3.797 & 0.49 & 0.55 & 1.09 
               & 83.8 $\pm$ 0.5  & $141.2^{+1.0}_{-1.0}$ 
               & 0.26            & 4.3 $\pm$ 2.1 
               & 16.7  $\pm$ 9.2 & 1.5 $\pm$ 0.3 & > 7.9
               & & O, md & 1, 2, 4; 5 -- 8\\ 

    SS\,Cyg     & UG      & 6.603 & 0.67 & 0.55 & 0.81
               & 51.0 $\pm$ 5.0  & 114.6$^{+0.6}_{-0.6}$
               & 0.70            & 25.4 $\pm$ 13.6
               & 36.2 $\pm$ 27.7 & 58.7 $\pm$ 11.8 & 13.5 -- 21.4
               & & O & 1 -- 3; 9, 10\\ 

    V347\,Pup   & NL      & 5.567  & 0.83 & 0.52 & 0.63
               & 84.0 $\pm$ 2.3   & 295.8$^{+1.4}_{-1.4}$
               & & 14.7 $\pm$ 7.5 & & & 
               & & & 1, 11, 12\\ 

    EX\,Dra     & UG      & 5.037 & 0.75 & 0.56 & 0.75
               & 84.2 $\pm$ 0.6  & 246.4$^{+1.2}_{-1.2}$
               & 0.43            & 10.7 $\pm$ 5.3 
               & 24.8 $\pm$ 14.3 & 3.5 $\pm$ 0.7 &
               & & O, d & 1, 2; 13 -- 15 \\ 
               
    U\,Gem      & UG      & 4.247 & 0.35 & 0.42 & 1.20
               & 69.7 $\pm$ 0.7  & 93.4$^{+0.3}_{-0.3}$
               & 0.32            & 6.2 $\pm$ 3.0 
               & 19.5 $\pm$ 10.7 & 6.4 $\pm$ 1.3 & 3.1 -- 4.0
               & & O, m & 1 -- 3; 16, 17  \\ 
               
    WZ\,Sge     & WZ      & 1.361 & 0.09 & 0.08 & 0.85
               & 77.0 $\pm$ 2.0  & 45.1$^{+0.1}_{-0.1}$
               & 0.04            & 0.2  $\pm$ 0.1 
               & 4.0 $\pm$ 2.7   & 0.04 $\pm$ 0.01&
               & & SO, rd & 1, 2; 18 -- 20 \\ 
               
    V3885\,Sgr  & NL      & 4.972  & 0.68 & 0.47 & 0.70
               & 65.0 $\pm$ 10.0  & 132.7$^{+1.4}_{-1.4}$
               & & 10.2 $\pm$ 5.1 & & & 39.6 -- 55.5
               & & & 1, 3; 21 \\ 
               
    DQ\,Her     & IP      & 4.647 & 0.62 & 0.4 & 0.6
               & 89.7 $\pm$ 0.1  & 500.6$^{+6.0}_{-5.9}$
               & 0.25            & 8.2 $\pm$ 4.1 
               & 22.1 $\pm$ 12.2 & 4.8 $\pm$ 1.0 &
               & & & 1; 22\\ 
               
    CRTS\,J1111 & SU      & 0.939 & 0.09 & 0.07 & 0.83
               & 71 $\pm$ 8      & $557^{+47}_{-40}$
               & 0.02            & 0.05 $\pm$ 0.04 
               & 2.4 $\pm$ 1.9   & 1.9 $\pm$ 0.5 & $\sim$ 3.0
               & & O & 23, 24 \\ 
               
    V406\,Vir   & WZ      & 1.342 & 0.07 & 0.07 & 1.00
               & 55 $\pm$ 5      & $169.2^{+4.6}_{-4.3}$
               & 0.04    & 0.2 $\pm$ 0.1 
               & 3.9 $\pm$ 2.7 & 1.9 $\pm$ 0.4 &
               & & Q & 1; 25, 26 \\ 
               
    EC21178-5417   & NL & 3.708  & 0.38 & 0.29 & 0.75
               & 75.0 $\pm$ 5.0  & 537.3$^{+9.1}_{-8.8}$
               & & 4.0 $\pm$ 2.0 & & &
               & & & 27 -- 29  \\ 
    \hline          
    V2779\,Oph  & NL      & 1.681 & 0.17 & 0.14 & 0.80
               & 77.1 $\pm$ 0.5  & 285.5$^{+6.0}_{-5.8}$
               & & 0.3 $\pm$ 0.2 & & &
               & & & 1; 30 -- 32 \\ 
               
    UX\,UMa     & UX      & 4.720 & 0.44 & 0.39 & 0.90
               & 70.0 $\pm$ 5.0 & 297.6$^{+2.1}_{-2.1}$
               & & 8.7 $\pm$ 4.3 & & & 15.9 -- 47.6
               & & & 1, 3; 33, 34\\ 
               
    HT\,Cas     & SU      & 1.768 & 0.15 & 0.09 & 0.61
               & 81.0 $\pm$ 1.0  & 141.4$^{+1.3}_{-1.2}$
               & 0.07            & 0.4 $\pm$ 0.2
               & 5.7 $\pm$ 3.5   & 4.5 $\pm$ 0.9 &
               & & Q & 1, 2; 35 \\ 
               
    V2051\,Oph  & SU      & 1.498 & 0.19 & 0.15 & 0.78
               & 83 $\pm$ 2  & 112.3$^{+0.9}_{-0.9}$
               & 0.05            & 0.2 $\pm$ 0.1 
               & 4.4 $\pm$ 2.9   & 2.7 $\pm$ 0.5 &
               & & Q & 1, 2; 36, 37\\ 
               
    CRTS\,J0359 & SU      & 1.910 & 0.28 & 0.17 & 0.60
               & 87 $\pm$ 2  & 445$^{+32}_{-28}$
               & 0.12            & 0.5 $\pm$ 0.3 
               & 4.0 $\pm$ 2.5   & 1.7 $\pm$ 0.4 &
               & & SO & 1, 2; 38\\ 
    
    V1838\,Aql  & WZ      & 1.369 & 0.09 & 0.08 & 0.89
               & 60 $\pm$ 5  & 202$^{+7}_{-7}$
               & 0.04            & 0.2 $\pm$ 0.1 
               & 4.0 $\pm$ 2.7   & 35.6 $\pm$ 7.6 &
               & & SO & 1, 2; 39\\ 
               
 	\hline \bottomrule
  \end{tabular}
  
  \begin{tablenotes}[para, flushleft]
  \footnotesize
  {\bf Notes:} \hspace{-0.5cm}
  \item[a] Classification of CVs: dwarf novae of the subtypes U\,Gem (UG), SU UMa (SU) and WZ\,Sge (WZ); novalikes of the subtypes UX\,UMa (UX), SW Sex (SW) or unknown subtype (NL); and intermediate polars (IP). \hspace{-0.5cm}
  \item[b] Distances derived from {\it Gaia} parallaxes (\citealt{2016A&A...595A...1G}, \citealt{2018A&A...616A...1G}, \citealt{2018A&A...616A...9L}). \hspace{-0.5cm}
  \item[c] Duty cycle estimated as a function of the orbital period \cite{2016MNRAS.456.4441C}. This expression is only valid for systems below the period gap. \hspace{-0.5cm}
  \item[d] Mass transfer rate from the secondary star estimated as a function of the orbital period \citep{1984ApJS...54..443P}. \hspace{-0.5cm}
  \item[e] Constraints on the mass accretion rates (see Section~\ref{ssec:acc_rate}). \hspace{-0.5cm}
  \item[f] Range of accretion rates published in the literature. \hspace{-0.5cm} 
  \item[g] Activity level when the spiral density waves were observed: outburst (O), superoutburst (SO) or quiescence (Q). The phase of the outburst is also indicated if known: rise (r), maximum (m) or decline (d). Note that this does not apply to novalikes.
  
  \vspace{0.1cm}
  \hfill\parbox[t]{22.92cm}{{\bf References: }
     [1] \citealt{2003A&A...404..301R}
     [2] The International variable Star Index Database (AAVSO)
     [3] \citealt{2018A&A...617A..26D} 
     [4] \citealt{2010MNRAS.402.1824C}
     [5] \citealt{1997MNRAS.290L..28S}
     [6] \citealt{1990ApJ...349..593M}
     [7] \citealt{1999MNRAS.306..348H}
     [8] \citealt{2000MNRAS.313..454M}
     [9] \citealt{1996MNRAS.281..626S}
     [10] \citealt{2012MmSAI..83..688K}
     [11] \citealt{1998MNRAS.299..545S}
     [12] \citealt{2005MNRAS.357..881T}     
     [13] \citealt{2000A&A...356L..33J}
     [14] \citealt{1996MNRAS.278..673B}
     [15] \citealt{2000A&A...354..579J}
     [16] \citealt{2001ApJ...551L..89G}
     [17] \citealt{1990ApJ...364..637M}
     [18] \citealt{2002PASJ...54L...7B}
     [19] \citealt{2004AN....325..185S}
     [20] \citealt{2000MNRAS.318..429S}
     [21] \citealt{2005MNRAS.363..285H}
     [22] \citealt{2010MNRAS.407.1903B}
     [23] \citealt{2013MNRAS.431..372C}
     [24] \citealt{2013AJ....145..145L}
     [25] \citealt{2010ApJ...711..389A}
     [26] \citealt{2019MNRAS.483.1080P}
     [27] \citealt{RuizCarmona_ec}
     [28] \citealt{Khangale:2013uq}
     [29] \citealt{Khangale2019}
     [30] \citealt{2008ATel.1640....1D}
     [31] \citealt{2011Obs...131...66S}
     [32] \citealt{2011AstL...37..845Y}
     [33] \citealt{2011MNRAS.410..963N}    
     [34] \citealt{1995ApJ...448..395B}
     [35] \citealt{2016A&A...586A..10N}
     [36] \citealt{2016MNRAS.463.3290R}
     [37] \citealt{2015MNRAS.447..149L}
     [38] \citealt{2018AJ....155..232L}
     [39] \citealt{2019MNRAS.486.2631H}
  }
 
  \end{tablenotes}

\end{threeparttable}
\end{table}
\end{landscape}


\subsubsection{Outburst phase}
\label{sssec:outb_phase}

The spiral structure that we can detect with Doppler tomography may  develop only at specific phases of the outburst cycle. We have detected and followed-up outbursts from very early stages, and PTFS\,1117an and V795\,Cyg were unambiguously observed during the rise of the outburst. We followed-up the systems for 4-5 days, falling short to fully cover the return to quiescence in many cases (see Table~\ref{tab:observations}). We note that the spiral structure in PTFS\,1200o and SV\,CMi is best visible at later phases relative to the outburst, so it is possible that spiral waves are more difficult to detect at the beginning of outbursts in some systems. On the other hand, the asymmetries present in some discs, e.g. V795\,Cyg, are stronger right after the peak of the outburst, fading at later phases without remarkable changes in position or extension. Thus, it seems unlikely that the systems that do not show spiral structure will do so when discs are closer to quiescence.


\section{Summary and Conclusions}
\label{sec:summary}

We have presented the first systematic search of spiral density waves with Doppler tomography in CVs, having solved the problem of the unpredictability of dwarf nova outbursts.

We find that spiral density waves are not always observed in Doppler tomograms of CVs in outburst and have a low occurrence rate. This does not mean that spiral density waves do not develop in the systems in which we do not detect them, but rather it is possible that spiral waves are not bright enough to be detected or that some conditions in the disc may complicate detection.

Comparing our sample with systems reported to have spiral structures we note the following characteristics that favour the detection of spiral structures: (i) high orbital inclination, and in particular systems that show pure emission lines during outbursts. (ii) the presence of He {\sc ii} $\lambda$4686 in emission.

As we also detected a number of systems that are at high inclination but do not show spiral structure but patchy phase-dependent emission, whose interpretation is uncertain, we strongly encourage the inclusion of radiative transfer calculations in hydrodynamical simulations of accretion discs to assess the influence of possible self-obscuration in spectral lines on the observability of underlying spiral structures. 

We suggest that future observational campaigns should aim for the following two targets. First, the number of systems for which spectroscopy in outburst is available needs to be expanded. With so many transient alert systems, this can be done on a nightly basis. Systems that show pure emission lines and He {\sc ii} emission should be considered for phase-resolved spectroscopy during outburst. We expect that spiral structures will be detectable in those systems. Second, follow-up campaigns for systems known to exhibit spiral waves are desirable. EX\,Dra is an illustrative example; since it recurs every three weeks and spends 60\% of the time in outburst, it provides an opportunity to test the full evolution of the accretion disc and the spiral structure throughout the outburst cycle. 


\section*{Acknowledgements}

This work is made possible by grant number 614.001.207 of the Netherlands Organisation for Scientific Research (NWO). RRC and PJG thank the Kavli Institute for Theoretical Physics at the University of California, Santa Barbara for valuable scientific exchangse during the programme DISKS17 from January to March 2017. This research was supported in part by the National Science Foundation under Grant No. NSF PHY17- 48958. RRC and PJG also thank the University of Cape Town for their hospitality, part of this work was done while visiting UCT supported by the NWO-NRF Bilateral Agreement in Astronomy.

The Palomar Transient Factory and the Intermediate Palomar Transient Factory project are a scientific collaboration among the California Institute of Technology, Los Alamos National Laboratory, the University of Wisconsin, Milwaukee, the Oskar Klein Center, the Weizmann Institute of Science, the TANGO Program of the University System of Taiwan, and the Kavli Institute for the Physics and Mathematics of the Universe. 

The William Herschel Telescope and the Isaac Newton Telescope are operated on the island of La Palma by the Isaac Newton Group of Telescopes in the Spanish Observatorio del Roque de los Muchachos of the Instituto de Astrof\'isica de Canarias. RRC thanks all staff of the Observatory for fun and productive visits. We thank Theo van Grunsven for observing PTFS\,1118m at the INT, and Thomas Wevers for observing V503\,Cyg at WHT. We also thank Cameron van Eck for helping with observations with the telescope at Rothney Astrophysical Observatory (University of Calgary, Canada). {\sl pt5m} is a collaborative effort between the Universities of Durham and Sheffield, kindly hosted at the WHT building. We thank Liam Hardy, Stuart Littlefield and Vik Dhillon for kindly scheduling our observations.

We acknowledge with thanks the variable star observations from the AAVSO International Database contributed by observers worldwide and used in this research. We thank Elizabeth Waagen and the many observers that participated in out monitoring campaigns. We have made extensive use of the International Variable Star Index website (\url{https://www.aavso.org/vsx/index.php}). We acknowledge use of services by the CV network (CVnet, \url{https://sites.google.com/site/aavsocvsection/}) and the Variable Star Network (VSNET, \url{http://www.kusastro.kyoto-u.ac.jp/vsnet/index.html}). This research has made use of the SIMBAD database, operated at CDS, Strasbourg, France.

This paper uses data from the Catalina Sky Survey and Catalina Real Time Survey; the CSS is funded by the National Aeronautics and Space Adminis- tration under Grant No. NNG05GF22G issued through the Science Mission Directorate Near-Earth Objects Observa- tions Program. The CRTS survey is supported by the U.S. National Science Foundation under grants AST-0909182.

Development of ASAS-SN has been supported by NSF grant AST-0908816 and the Center for Cosmology and AstroParticle Physics at The Ohio State University.

The ASAS project was supported to a large extent by the generous donation of William Golden. This work was partially supported by the Polish MNiI grants no. 1P03D-008-27 and 2P03D-020-24.

The Pan-STARRS1 Surveys (PS1) have been made possible through contributions of the Institute for Astronomy, the University of Hawaii, the Pan-STARRS Project Office, the Max-Planck Society and its participating institutes, the Max Planck Institute for Astronomy, Heidelberg and the Max Planck Institute for Extraterrestrial Physics, Garching, The Johns Hopkins University, Durham University, the University of Edinburgh, Queen's University Belfast, the Harvard-Smithsonian Center for Astrophysics, the Las Cumbres Observatory Global Telescope Network Incorporated, the National Central University of Taiwan, the Space Telescope Science Institute, the National Aeronautics and Space Administration under Grant No. NNX08AR22G issued through the Planetary Science Division of the NASA Science Mission Directorate, the National Science Foundation under Grant No. AST-1238877, the University of Maryland, and Eotvos Lorand University (ELTE). 

The SDSS is managed by the Astrophysical Research Consortium for the Participating Institutions. Funding for the Sloan Digital Sky Survey IV has been provided by the Alfred P. Sloan Foundation, the U.S. Department of Energy Office of Science, and the Participating Institutions. SDSS-IV acknowledges support and resources from the Center for High-Performance Computing at the University of Utah. The SDSS web site is \url{http://www.sdss.org/}. 

This work has made use of data from the European Space Agency (ESA) mission {\it Gaia} (\url{https://www.cosmos.esa.int/gaia}), processed by the {\it Gaia} Data Processing and Analysis Consortium (DPAC, \url{https://www.cosmos.esa.int/web/gaia/dpac/consortium}). Funding for the DPAC has been provided by national institutions, in particular the institutions participating in the {\it Gaia} Multilateral Agreement.

Based on observations made with the NASA Galaxy Evolution Explorer. GALEX is operated for NASA by the California Institute of Technology under NASA contract NAS5-98034. 

{\sc iraf} is distributed by the National Optical Astronomy Observatories, which are operated by the Association of Universities for Research in Astronomy, Inc., under cooperative agreement with the National Science Foundation \citep{1993ASPC...52..173T}. {\sc pyraf} and {\sc pyfits} are products of the Space Telescope Science Institute, which is operated by AURA for NASA. 

We have made extensive use of {\sc python} packages: {\sc numpy} \citep{Oliphant:2006wm}, {\sc matplotlib} \citep{Hunter:2007}, {\sc scipy} \citep{2001..scipy}, and {\sc astropy} \citep{2013A&A...558A..33A}. We have also use the package {\sc starlink-pyndf} by T. Jenesses, G. Bell and T. Marsh, which is publicly available on the website \url{https://github.com/timj/starlink-pyndf}. 

We have also used the software {\sc molly} and {\sc doppler} by T. Marsh for preparing spectra and computing Doppler tomograms. This software is publicly available on the website \url{http://deneb.astro.warwick.ac.uk/phsaap/software/}.




\bibliographystyle{mnras}
\bibliography{sdw_observations}


\appendix


\section{Photometric summary of observed systems}
\label{sec:photometry}

\subsection{Long-term photometric monitoring}
\label{sec:longterm_lc}

In this Section, we collect and present the photometric data available from a wealth of observational facilities. We include the Palomar Transient Factory (PTF; \citealt{2009PASP..121.1334R}, \citealt{2009PASP..121.1395L}), the Catalina Real Time Survey (CRTS; \citealt{2009ApJ...696..870D}), the All-Sky Automated Survey for Supernovae (ASAS-SN; \citealt{2017PASP..129j4502K}), the All-Sky Automated Survey (ASAS; \citealt{1997AcA....47..467P}), the American Association of Variable Star Observers (AAVSO).

\begin{figure*}
  \caption{Photometric monitoring of observed systems. Horizontal dotted lines indicate detection limits and vertical lines indicate the time of our spectroscopic observations.}
  \label{fig:monitoring1}
  \centering
    \includegraphics[width=0.95\textwidth]{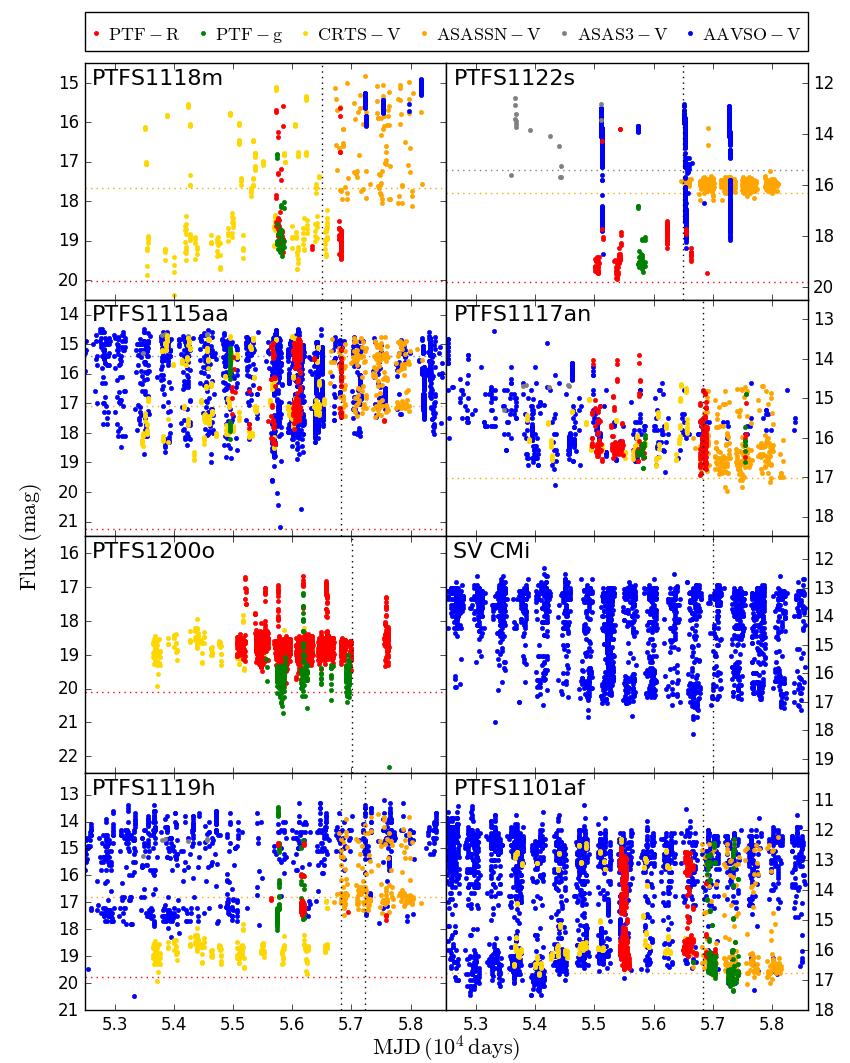}
\end{figure*}

\begin{figure*}
  \caption{Photometric monitoring of observed systems (continuation). Horizontal dotted lines indicate detection limits and vertical lines indicate the time of our spectroscopic observations}
  \label{fig:monitoring2}
  \centering
    \includegraphics[width=0.95\textwidth]{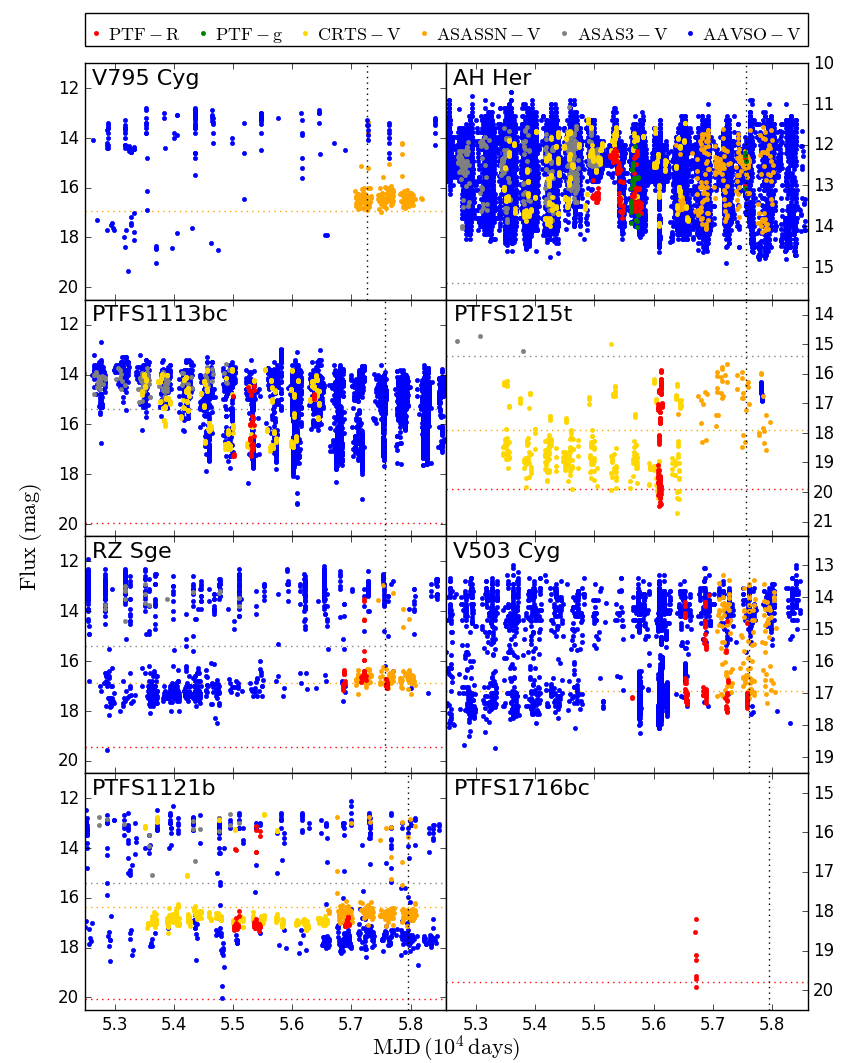}
\end{figure*}

We display the long-term light curves in Figures~\ref{fig:monitoring1} and \ref{fig:monitoring2}, from 2002 until the present day. We obtain representative statistical values from the light curves, such as the median magnitude of the system with the corresponding scatter. We also calculate the quiescent magnitude by averaging the data points that are not in outburst. We estimate the outburst magnitude as well, taken as the maximum value registered in the light curve. This is a lower limit, since the peak luminosity of the outburst could be missed in the observations. These values are gathered in Table~\ref{tab:longtermlc_data}.

In addition, we present the reported magnitudes as observed by the Panoramic Survey Telescope and Rapid Response System (PanSTARRS PS1; \cite{2010SPIE.7733E..0EK}) and Sloan Digital Sky Survey (SDSS-DR15; \citealt{2019ApJS..240...23A}) in Table~\ref{tab:r_and_uv}. PanSTARRS provides the RMS scatter of the multiple detections, which is very helpful in evaluating the observed level of activity. We use both measurements to approximate the quiescent magnitude of the systems in the $r$-filter. Furthermore, we present the UV flux observed by the NASA Galaxy Evolution Explorer (GALEX) in the available bands. 


\onecolumn
\begin{center}
\renewcommand*{\arraystretch}{1.2}
\begin{longtable}[c]{lcccrr cl} 

  \captionsetup{width=0.9\textwidth}
  \caption{Summary of the information extracted from available light curves obtained by long-term monitoring surveys. Included are median magnitudes, quiescent and outburst magnitudes, number of detections and number of observed outbursts. The asterisk (*) indicates more than 30 outbursts.}
  \label{tab:longtermlc_data}
  
    \\ \hline \hline
    System  & $m_{\rm{median}}\; [\sigma_{\rm{RMS}}]$  
            & $m_{\rm{quiesc}}\; [\sigma_{\rm{RMS}}]$  
            & $m_{\rm{outb}}$  & $N_{\rm{det}}$   & $N_{\rm{outb}}$   & Filter    & Survey \\
    \hline
    \endhead
    
    \hline \multicolumn{8}{r}{{Continued on next page}} \\ 
    \endfoot

    \hline \hline
    \endlastfoot
    
    PTFS\,1118m  & 18.86 $\pm$ 0.04 [1.00] & 19.10 $\pm$ 0.04 [0.18] 
               & 15.58 $\pm$ 0.01        &    84 &  6 & R & PTF \\
               & 18.57 $\pm$ 0.14 [1.28] & 19.06 $\pm$ 0.14 [0.32] 
               & 15.10 $\pm$ 0.06        &   240 & 14 & V & CRTS \\
               & 16.60 $\pm$ 0.15 [0.99] & -- 
               & 14.84 $\pm$ 0.05        &   103 & 20 & V & ASAS-SN \\
               & 15.46 $\pm$ 0.03 [0.17] & -- 
               & 14.90 $\pm$ 0.05        &   1510 & 4 & V & AAVSO \\               
    PTFS\,1122s  & 18.69 $\pm$ 0.05 [0.96] & 19.10 $\pm$ 0.05 [0.27] 
               & 13.79 $\pm$ 0.01        &   125 &  2 & R & PTF \\
               & 15.98 $\pm$ 0.11 [0.53] & -- 
               & 13.03 $\pm$ 0.02        &   384 &  3 & V & ASAS-SN \\
               & 13.65 $\pm$ 0.07 [1.02] & -- 
               & 12.80 $\pm$ 0.07        &    16 &  5 & V & ASAS-3 \\
               & 13.84 $\pm$ 0.15 [1.05] & -- 
               & 12.80 $\pm$ 0.05        &  9156 &  4 & V & AAVSO \\
               
 	PTFS\,1115aa & 16.14 $\pm$ 0.07 [1.02] & 17.21 $\pm$ 0.07 [0.52]
               & 14.82 $\pm$ 0.00        &   239 & 18 & R & PTFS \\
               & 16.84 $\pm$ 0.08 [1.18] & 17.53 $\pm$ 0.08 [0.48]
               & 14.70 $\pm$ 0.06        &   344 & 18 & V & CRTS \\
               & 15.75 $\pm$ 0.11 [0.82] & --
               & 14.75 $\pm$ 0.04        &   184 &  * & V & ASAS-SN \\
               & 14.69 $\pm$ 0.05 [0.23] & --
               & 14.66 $\pm$ 0.05        &     6 &  4 & V & ASAS-3 \\
               & 15.64 $\pm$ 0.05 [0.95] & --
               & 14.10 $\pm$ 0.05        &  8074 &  * & V & AAVSO \\
               
 	PTFS\,1117an & 16.17 $\pm$ 0.01 [0.57] & 16.38 $\pm$ 0.01 [0.17] 
               & 13.90 $\pm$ 0.01        &   229 & 13 & R & PTFS\\
               & 16.01 $\pm$ 0.07 [0.53] & 16.29 $\pm$ 0.07 [0.18] 
               & 14.63 $\pm$ 0.06        &   119 &  9 & V & CRTS\\
               & 16.36 $\pm$ 0.11 [0.60] & 16.68 $\pm$ 0.11 [0.21] 
               & 14.68 $\pm$ 0.04        &   226 &  * & V & ASAS-SN\\
               & 14.65 $\pm$ 0.03 [1.14] & --
               & 10.96 $\pm$ 0.10        &  1661 &  * & V & AAVSO\\
               
 	PTFS\,1200o  & 18.70 $\pm$ 0.05 [0.33] & 18.89 $\pm$ 0.05 [0.16] 
               & 16.68 $\pm$ 0.01        &  2619 &  6 & R & PTF\\
               & 18.74 $\pm$ 0.16 [0.45] & 19.02 $\pm$ 0.16 [0.23] 
               & 16.94 $\pm$ 0.08        &   211 &  3 & V & CRTS\\
               
 	SV\,CMi     & 13.60 $\pm$ 0.04 [1.04] & 17.19 $\pm$ 0.05 [0.24]
               & 12.00 $\pm$ 0.10        &  6285 &  * & V & AAVSO\\
               
 	PTFS\,1119h  & 17.17 $\pm$ 0.01 [0.88] & 17.34 $\pm$ 0.01 [0.11] 
               & 14.80 $\pm$ 0.02        &    54 &  4 & R & PTFS\\
               & 16.66 $\pm$ 0.12 [1.01] & 16.93 $\pm$ 0.12 [0.18]
               & 13.71 $\pm$ 0.03        &   165 &  * & V & ASAS-SN\\
               & 15.20 $\pm$ 0.04 [1.03] & --
               & 13.20 $\pm$ 0.10        &  1620 &  * & V & AAVSO\\
               
 	PTFS\,1101af & 15.72 $\pm$ 0.01 [1.33] & 16.03 $\pm$ 0.01 [0.25] 
               & 12.59 $\pm$ 0.00        &   991 & 13 & R & PTFS\\
               & 15.81 $\pm$ 0.06 [1.42] & 16.11 $\pm$ 0.06 [0.23] 
               & 12.29 $\pm$ 0.05        &   271 & 17 & V & CRTS\\
               & 16.35 $\pm$ 0.12 [1.47] & 16.62 $\pm$ 0.12 [0.16]
               & 12.23 $\pm$ 0.01        &   220 &  * & V & ASAS-SN\\
               & 12.85 $\pm$ 0.05 [1.48] & --
               & 10.80 $\pm$ 0.10        &  8481 &  * & V & AAVSO\\

 	V795\,Cyg   & 18.19 $\pm$ 0.12 [0.30] & 18.40 $\pm$ 0.12 [0.11] 
               & 17.70 $\pm$ 0.06        &     4 &  1 & R & PTFS\\
               & 16.42 $\pm$ 0.15 [0.54] & --
               & 13.36 $\pm$ 0.02        &   156 &  4 & V & ASAS-SN\\
               & 13.21 $\pm$ 0.02 [1.25] & 17.93 $\pm$ 0.10 [0.54]
               &  8.60 $\pm$ 0.10        &   449 &  * & V & AAVSO\\
               
 	AH\,Her     & 12.90 $\pm$ 0.01 [0.51] & 13.33 $\pm$ 0.01 [0.20] 
               & 12.11 $\pm$ 0.01        &   123 &  * & R & PTFS\\
               & 12.48 $\pm$ 0.05 [0.70] & 13.22 $\pm$ 0.05 [0.44] 
               & 11.37 $\pm$ 0.05        &   422 &  * & V & CRTS\\
               & 12.54 $\pm$ 0.05 [0.67] & 13.31 $\pm$ 0.05 [0.47]
               & 11.57 $\pm$ 0.02        &   229 &  * & V & ASAS-SN\\
               & 12.43 $\pm$ 0.04 [0.53] & 12.92 $\pm$ 0.04 [0.43]
               & 11.08 $\pm$ 0.05        &   348 &  * & V & ASAS-3\\
               & 12.48 $\pm$ 0.02 [0.91] & 13.35 $\pm$ 0.02 [0.62]
               & 10.50 $\pm$ 0.05        & 50679 &  * & V & AAVSO\\
               
 	PTFS\,1113bc & 16.00 $\pm$ 0.01 [1.03] & 16.80 $\pm$ 0.01 [0.41] 
               & 14.45 $\pm$ 0.01        &    37 &  5 & R & PTFS\\
               & 14.96 $\pm$ 0.06 [0.99] & 16.06 $\pm$ 0.06 [0.68] 
               & 13.76 $\pm$ 0.05        &   342 & 20 & V & CRTS\\
               & 14.25 $\pm$ 0.06 [0.30] & --
               & 13.56 $\pm$ 0.07        &   105 & 14 & V & ASAS-3\\
               & 14.73 $\pm$ 0.03 [0.94] & --
               & 12.70 $\pm$ 0.10        & 51215 &  * & V & AAVSO\\
    
 	PTFS\,1215t  & 19.47 $\pm$ 0.07 [1.52] & 19.84 $\pm$ 0.07 [0.26] 
               & 15.88 $\pm$ 0.01        &   133 &  4 & R & PTFS\\
               & 18.75 $\pm$ 0.16 [1.01] & 19.21 $\pm$ 0.16 [0.36] 
               & 14.98 $\pm$ 0.06        &   349 & 14 & V & CRTS\\
               & 16.76 $\pm$ 0.14 [0.81] & --
               & 15.69 $\pm$ 0.09        &    53 &  6 & V & ASAS-SN\\
               & 14.91 $\pm$ 0.04 [0.21] & --
               & 14.71 $\pm$ 0.04        &     3 &  1 & V & ASAS-3\\
               & 16.52 $\pm$ 0.00 [0.15] & --
               & 16.29 $\pm$ 0.00        &   383 &  1 & V & AAVSO\\
 	
 	RZ\,Sge     & 16.75 $\pm$ 0.02 [0.83] & 16.97 $\pm$ 0.02 [0.10] 
               & 13.52 $\pm$ 0.01        &    50 &  1 & R & PTFS\\
               & 16.69 $\pm$ 0.13 [0.72] & 16.85 $\pm$ 0.13 [0.14]
               & 12.95 $\pm$ 0.02        &   115 &  6 & V & ASAS-SN\\
               & 13.45 $\pm$ 0.04 [0.37] & --
               & 12.90 $\pm$ 0.04        &    17 &  7 & V & ASAS-3\\
               & 13.20 $\pm$ 0.03 [1.60] & 17.32 $\pm$ 0.10 [0.26]
               & 11.30 $\pm$ 0.10        &  3098 &  * & V & AAVSO\\ 
               
 	V503\,Cyg   & 17.13 $\pm$ 0.02 [1.01] & 17.28 $\pm$ 0.02 [0.09] 
               & 13.94 $\pm$ 0.01        &   166 &  6 & R & PTFS\\
               & 15.18 $\pm$ 0.13 [1.13] & 16.37 $\pm$ 0.13 [0.57]
               & 13.30 $\pm$ 0.03        &   131 & 16 & V & ASAS-SN\\
               & 14.80 $\pm$ 0.04 [1.34] & 17.33 $\pm$ 0.05 [0.26]
               & 12.90 $\pm$ 0.05        & 16725 &  * & V & AAVSO\\

 	PTFS\,1121b  & 17.07 $\pm$ 0.01 [0.84] & 17.15 $\pm$ 0.01 [0.05] 
               & 13.14 $\pm$ 0.01        &   122 &  3 & R & PTFS\\
               & 16.89 $\pm$ 0.08 [1.00] & 17.00 $\pm$ 0.08 [0.09] 
               & 12.65 $\pm$ 0.05        &   351 &  7 & V & CRTS\\
               & 16.61 $\pm$ 0.13 [0.83] & 16.75 $\pm$ 0.13 [0.12]
               & 12.77 $\pm$ 0.02        &   217 & 10 & V & ASAS-SN\\
               & 13.07 $\pm$ 0.05 [0.67] & --
               & 12.66 $\pm$ 0.05        &    19 &  6 & V & ASAS-3\\
               & 13.50 $\pm$ 0.09 [1.88] & --
               & 11.90 $\pm$ 0.10        &  1578 & 30 & V & AAVSO\\

 	PTFS\,1716bc & 19.25 $\pm$ 0.09 [0.59] & 19.63 $\pm$ 0.09 [0.24] 
               & 18.20 $\pm$ 0.16        &     7 &  0 & R & PTFS\\
               
 	\hline

\end{longtable}
\end{center}
\twocolumn

\onecolumn
\begin{table*}
  \centering
  \renewcommand*{\arraystretch}{1.2}
  \caption{Additional photometric information on the observed systems, including $r$-band magnitudes from PanSTARRS (including RMS scatter and number of detections) and SDSS (including estimates of the extinction). The UV flux as reported by GALEX is included where available.}
  \label{tab:r_and_uv}
  \begin{tabular}{lcr c cc c 
                 @{\hskip 0.2in} S[table-format=0.1]
                 @{\hskip 0.3in} S[table-format=0.1]} 
 	
    \hline
    \multirow{2}{1.2cm}{\centering{System}} & 
    \multicolumn{2}{c}{PanSTARRS} &&
    \multicolumn{2}{c}{SDSS} &&
    \multicolumn{2}{c}{GALEX [$\mu\rm{Jy}$]}\\
    
    \cline{2-3}
    \cline{5-6}
    \cline{8-9}

    & $r\; [\sigma_{\rm{RMS}}]$ & $N_{\rm{det}}$ & & $r$ & $A_r$ &
    & \hspace{0.8cm} FUV \hspace{0.4cm} & \hspace{0.5cm} NUV \hspace{2cm}\\
    \hline
    
    PTFS\,1118m  & 18.74 $\pm$ 0.02 [0.21] & 12 &
               & 19.37 $\pm$ 0.02 & 0.08 &
               & 34.9  $\,\pm\,$ 5.2  & 56.8 $\,\pm\,$ 5.6 \\
    PTFS\,1122s  & 18.61 $\pm$ 0.09 [0.32] & 14 & 
               & 19.34 $\pm$ 0.01 & 0.23 &
               & 39.6  $\,\pm\,$ 4.9  & 49.2 $\,\pm\,$ 3.9 \\
 	PTFS\,1115aa & 15.41 $\pm$ 0.03 [0.18] & 12 &
        	   & 16.18 $\pm$ 0.00 & 0.09 & \\
 	PTFS\,1117an & 15.85 $\pm$ 0.00 [0.01]& 12 &
 	           & 14.24 $\pm$ 0.00 & 0.06 & 
 	           & 357.3 $\,\pm\,$ 4.9 &\\
 	PTFS\,1200o  & 18.72 $\pm$ 0.03 [0.11] & 14 & 
 	           & 17.77 $\pm$ 0.01 & 0.14 & 
 	           & & 9.9 $\,\pm\,$ 5.6  \\
 	SV\,CMi     & -- & 0 & \\
 	PTFS\,1119h  & 16.76 $\pm$ 0.21 [0.77]& 18 &
 	           & & & 
 	           & & 258.3 $\,\pm\,$ 3.9  \\
 	PTFS\,1101af & 14.77 $\pm$ 0.16 [0.86] & 18 &
 	           & & & 
 	           & & 727.7 $\,\pm\,$ 19.8  \\
 	V0795 Cyg  & 18.39 $\pm$ 0.01 [0.02] & 9 &
 	           & & & 
 	           & & \\
 	AH\,Her     & -- & 0 &
 	           & 14.65 $\pm$ 0.01 & 0.10 &  
 	           & 12515.7 $\,\pm\,$ 17.6 & 14321.7 $\,\pm\,$ 6.0 \\
 	PTFS\,1113bc & 14.41 $\pm$ 0.13 [1.21] & 11 &
 	           & 15.73 $\pm$ 0.00 & 0.05 & 
 	           & 4361.7 $\,\pm\,$ 63.4 & 5357.6 $\,\pm\,$ 45.9 \\
 	PTFS\,1215t  & 18.87 $\pm$ 0.11 [0.42] & 10 &
 	           & 18.47 $\pm$ 0.01 & 0.09 & 
 	           & 139.7 $\,\pm\,$ 9.0 & 173.3 $\,\pm\,$ 6.2  \\
 	RZ\,Sge     & 17.10 $\pm$ 0.02 [0.06] & 14 &
 	           & & & 
 	           & & 371.2 $\,\pm\,$ 10.8 \\
 	V0503 Cyg  & 16.69 $\pm$ 0.09 [0.40] & 16 &
 	           & & & 
 	           & & 257.6 $\,\pm\,$ 12.2 \\
 	PTFS\,1121b  & 17.86 $\pm$ 0.01 [0.03] & 8 &
 	           & 17.45 $\pm$ 0.01 &  0.06 &
 	           & 144.4 $\,\pm\,$ 10.3 & 139.5 $\,\pm\,$ 6.0 \\
 	PTFS\,1716bc & 19.13 $\pm$ 0.12 [0.73] & 16 &
 	           & & & 
 	           & 55.0 $\,\pm\,$ 8.1 & 53.3 $\,\pm\,$ 4.9  \\
 	
 	\hline
  \end{tabular}
\end{table*}
\twocolumn

\subsection{Light curve fitting}
\label{ssec:lcfitting}

We studied the light curves of three systems in order to obtain their orbital periods. 

\begin{figure}
  \caption{Light curve of PTFS\,1118m}
  \label{fig:lc_ptfs1118m}
  \centering
    \includegraphics[width=0.47\textwidth]{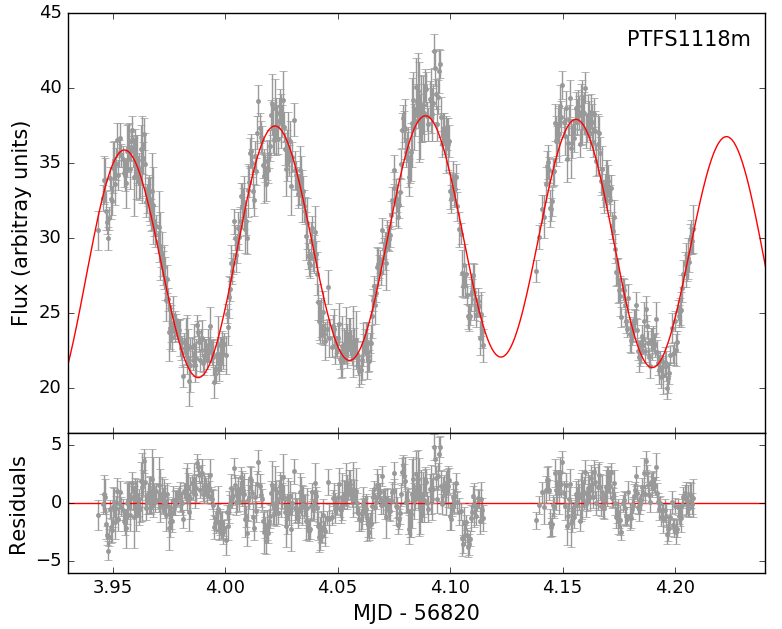}
\end{figure}

PTFS\,1118m was observed photometrically with the Wide Field Camera mounted on the Isaac Newton Telescope (La Palma), and the light curve, extracted with aperture photometry, is presented in Figure~\ref{fig:lc_ptfs1118m}. We fitted a sinusoidal function plus a polynomial component to account for the overall trend, and obtained an orbital period of $P_{\rm{orb}} = 1.6281 \pm 0.0026$ hours.

\begin{figure}
  \caption{Light curve of PTFS\,1200o}
  \label{fig:lc_ptfs1200o}
  \centering
    \includegraphics[width=0.47\textwidth]{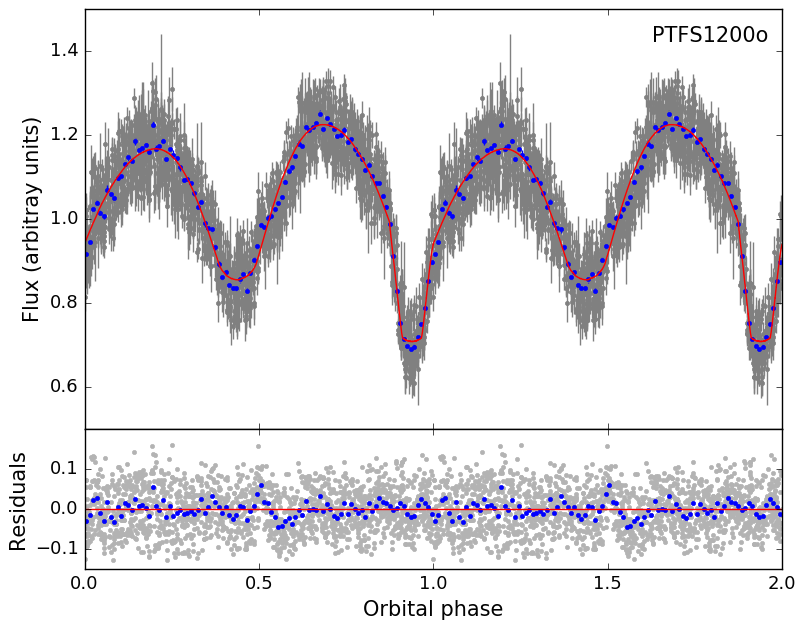}
\end{figure}

PTFS\,1200o has been extensively observed with the Palomar Transient Factory, as shown in Figure~\ref{fig:lc_ptfs1200o}. We use the observations in the Mould $R$-filter, and exclude data points at magnitudes lower than $R < 18.3$, i.e. during outburst. We further cleaned the light curve applying a sigma-clipping algorithm to the phase folded light curve, using a tentative orbital period of 6.28 hours. We fit the remaining data points in the light curve with \textsc{lcurve}, a code by T. R. Marsh and collaborators (\citealt{2010MNRAS.402.1824C} for an example). The free parameters of our model are the orbital period, the mid-time of primary eclipse, the inclination and the mass ratio, while other parameters such as the temperature of the stellar components were fixed to typical values. We obtained an orbital period $P_{\rm{orb}} = 6.27846 \pm 0.00002$ hours. The resulting ephemeris is:
\begin{equation}
  \label{eq:eph_ptfs1200o}
  \rm{HJD} = (2456500.484053 \pm 10^{-6}) + (0.261602 \pm 10^{-6}) E.
\end{equation}

The inclination of the system is determined to be $i=84^{\circ} \pm 2^{\circ}$, and the mass ratio to be in the range 0.31 < $q$ < 0.70.

\begin{figure}
  \caption{Light curve of PTFS\,1716bc}
  \label{fig:lc_ptfs1716bc}
  \centering
    \includegraphics[width=0.47\textwidth]{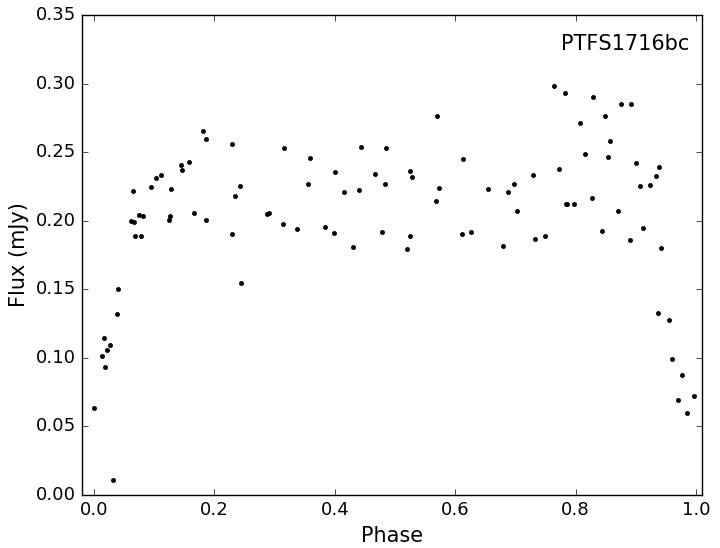}
\end{figure}

For PTFS\,1716bc, we extracted the light curve by integrating the continuum level in spectra taken with the ISIS red arm over wavelength ranges that are free of any spectral lines. Since the brightness of the system was declining towards quiescence, we applied a linear correction of 0.05 mJy day$^{-1}$ in order to bring all data points to the same flux level. We found that PTFS\,1716bc is an eclipsing system, and we report an orbital period $P_{\rm{orb}} = 2.387 \pm 0.010$ hours. We present the phase-folded light curve in Figure~\ref{fig:lc_ptfs1716bc}.



\section{Properties of the systems}
\label{sec:calc_pars}

\subsection{Orbital period and superhumps}

To estimate the orbital period and the mass ratio from the superhump period excess, $\varepsilon = P_{\rm{SH}} / P_{\rm{orb}} - 1$, where $P_{\rm{SH}}$ is the superhump period and $P_{\rm{orb}}$ is the orbital period, we used the following expressions (\citealt{2009MNRAS.397.2170G}, \citealt{2005PASP..117.1204P}):
\begin{equation}
    \label{eq:psh-porb}
    P_{\rm{orb}} = 0.9162(52)\, P_{\rm{SH}} + 5.39(52)\;  \rm{[min]} 
\end{equation}
\begin{equation}
    \label{eq:eps-q}
    \varepsilon(q) = 0.18 q + 0.29 q^2 \;(q < 0.38)  
\end{equation}

\cite{2013PASJ...65..115K} proposed a refinement of the estimate of the mass ratio, using the superhump period at the growing stage. However, since the stage of the reported values of superhump periods is often unclear, we prefer to use expression~\ref{eq:eps-q}.

\subsection{Mass-radius relation for the secondary star}

The secondary stars in CVs fill their Roche lobes, providing a relation between their mass and radii \citep{1972ApJ...175L..79F}:
\begin{equation}
    \label{eq:faulkner_sec}
    P_{\rm{orb}}\, [\rm{h}] = 8.75\, (m_2/r_2^3)^{-1/2}\;  \rm{[\odot]}
\end{equation}

\cite{2011ApJS..194...28K} fitted the empirical mass-radius relation with the following broken-power-law:
 
 \begin{equation}
    \label{eq:knigge_sec}
    R_2\, [\odot] = \left\{ \begin{array}{lcc}
    
    0.118 \pm 0.004 \left( {M_2} / {M_{\rm{bounce}}} \right)^{0.30 \pm 0.03} \\ \quad \rm{if}\; M_2 < M_{\rm{bounce}} \\ \\
        
    0.225 \pm 0.008 \left( {M_2} / {M_{\rm{conv}}} \right)^{0.61 \pm 0.01} \\ \quad \rm{if}\; M_{\rm{bounce}} < M_2 < M_{\rm{conv}} \\ \quad \rm{and}\; P_{\rm{min}} < P_{\rm{orb}} < P_{\rm{gap,-}}\\ \\
        
    0.293 \pm 0.010 \left( {M_2} / {M_{\rm{conv}}} \right)^{0.69 \pm 0.03} \\ \quad \rm{if}\; M_{\rm{conv}} < M_2 < M_{\rm{evol}}
    \\ \quad \rm{and}\; P_{\rm{gap,+}} < P_{\rm{orb}} < P_{\rm{evol}}\\
                  
        \end{array}
        \right.
\end{equation}

with $M_{\rm{bounce}} = 0.069 \pm 0.009 M_{\rm{\odot}}$, $M_{\rm{conv}} = 0.20 \pm 0.02 M_{\rm{\odot}}$, $M_{\rm{evol}} \simeq 0.6 - 0.7 M_{\rm{\odot}}$, $P_{\rm{min}} = 82.4 \pm 0.7$ min, $P_{\rm{gap,-}} = 2.15 \pm 0.03$ h, $P_{\rm{gap,+}} = 3.18 \pm 0.04$ h, $P_{\rm{evol}} \simeq 5-6$ h

For a given orbital period, expressions \ref{eq:faulkner_sec} and \ref{eq:knigge_sec} intersect which we use to disentangle mass and radius.

\subsection{Absolute magnitude of dwarf-nova outburst and inclination}

The absolute magnitude of dwarf-nova outbursts, $V_{\rm{outb}}$, is empirically related to the orbital period according to the expression (\citealt{2011MNRAS.411.2695P}): 
\begin{equation}
    \label{eq:Vout-porb}
    V_{\rm{outb}} = 4.95 - 0.199 P_{\rm{orb}}\,
\end{equation}
with the $V_{\rm outb}$ in magnitudes and the orbital period $P_{\rm orb}$ in hours and 
where no distinction is made between normal and superoutbursts. The relation has a root-mean-square scatter of 0.41 mag.

By comparison with the apparent magnitude, $v_{\rm{outb}}$, the distance of the system can be estimated:
\begin{equation}
    \label{eq:pogson}
    v_{\rm{outb}} - V_{\rm{outb}} - \Delta V_{i} = 5 \log d - 5
\end{equation}

where $\Delta V_{i}$ is a correction of the absolute magnitude that depends on the inclination $i$ of a flat, limb-darkened accretion disc \citep{1980AcA....30..127P}:
\begin{equation}
    \label{eq:incl_correction}
    \Delta V_{i} = -2.5 \log [ (1 + 3/2 \cos i) \cos i]
\end{equation}

In this work, we calculate the distances from \textit{Gaia} parallaxes. Therefore, we can use the absolute magnitude of outburst to estimate the inclination of the system. If the distance is not known, an average value of $i = 56.7^{\circ}$ is assumed for the inclination \citep{2016MNRAS.456.4441C}.

\subsection{Mass-accretion rate}
\label{ssec:acc_rate}

\subsubsection{From mass-transfer rate}

The secular mass transfer rate from the secondary, $\dot{M_2}$, scales with the orbital period according to the following expression \citep{1984ApJS...54..443P}:
\begin{equation}
    \label{eq:m2dot-porb}
    \dot{M_2} = (5.1 \pm 2.5) 10^{-10}\, (P_{\rm{orb}}/ 4 h)^{3.2 \pm 0.2}\,[M_{\rm{\odot}}\, \rm{yr}^{-1}],
\end{equation}
which works well for shorter orbital periods.

In steady-state accretion and without any other sources of mass loss (e.g. disk winds), the transfer rate from the secondary will equal the secular mass accretion rate onto the primary, $\dot{M_1}$. If we assume mass accretion only occurs during outbursts at a constant pace, the instantaneous mass accretion rate onto the white dwarf $\dot{M_1}_{\rm{,}\,dc}$ can be estimated if we know how much time a system spends in outburst, i.e. the duty cycle.
\begin{equation}
    \dot{M_1}_{\rm{,}\,dc} \approx \, \dot{M_2} / dc
\end{equation}

Following \citep{2016MNRAS.456.4441C} the duty cycle can be estimated from:
\begin{equation}
    \label{eq:dc}
    \log(dc) = (1.9 \pm 0.4)\, \log(P_{\rm{orb}} [h]) - (1.63 \pm 0.08)
\end{equation}
This is only valid for systems below the period gap, but we extrapolated this relation for the rest of the systems.

This accretion rate during outburst is an upper limit to the secular accretion rate onto the white dwarf, since it neglects the possibility of accretion during quiescence. On the other hand we also omit any mass loss in e.g. winds would decrease the effective accretion rate onto the white dwarf. 

\subsubsection{From outburst luminosity in the optical}

Alternatively, we can use the observed maximum and minimum magnitudes in the V-band, $v_{\rm{quies}}$ and $v_{\rm{outb}}$, converted into flux, to estimate the mass-accretion rate during outburst. 

We argue that the excess flux in outburst in the optical only comes from the disc. Neglecting any bolometric corrections, the luminosity during outburst is:
\begin{equation}
    \label{eq:lum_outb}
    L_{\rm{acc}} = 4 \pi d^2 (F_{\rm{outb}} - F_{\rm{quiesc}})
\end{equation}

We can estimate the mass-accretion rate from the accretion luminosity assuming that the gas in the disc releases half of its potential energy in the optical during accretion:
\begin{equation}
    \label{eq:lum_acc}
    \dot{M_1}_{\rm{,}\,out} = \frac{2 L_{\rm acc} R_1}{G M_1}
\end{equation}

where $R_1$ is estimated from  $M_1$ according to the mass-radius relation for white dwarfs by \cite{1988ApJ...332..193V}. 
As we are estimating the outburst flux solely from optical flux, this will be a lower limit to the instantaneous accretion rate, as we are neglecting emission at other wavelengths. Moreover, the flux in outburst may be understimated since the peak of the outburst may have not been observed. On the other hand, the flux in outburst most likely correspond to superoutburst for the systems below the period gap; but this is typically a small deviation from normal outburst.


\section{Comments on the detection of spiral density waves in the literature}
\label{sec:comm_sdws}

We briefly review the occurrence of spiral density waves published in the literature:

\begin{itemize}
    \item IP\,Peg: the first system that was reported to show strong spiral density waves in its disc \citep{1997MNRAS.290L..28S}. Spiral structure has been detected in several outbursts since then, especially clear in He {\sc ii}$~\lambda$4686 (\citealt{1999MNRAS.306..348H}, \citealt{2000MNRAS.313..454M}). There is less robust evidence of spiral structure during outburst in tomograms published in \citealt{1990ApJ...349..593M}, \citealt{1996MNRAS.281..626S}).
    
    \item SS\,Cyg showed extremely faint stationary emission regions at the expected phases of spiral density waves during outburst \citep{1996MNRAS.281..626S}. In quiescence, the Balmer tomograms show some spots of enhanced emission towards the lower right part of the tomogram \citep{2012MmSAI..83..688K}.

    \item V347\,Pup showed strong spiral waves in H$\beta$ \citep{1998MNRAS.299..545S}, that were clearer in He {\sc i} lines at a later epoch \citep{2005MNRAS.357..881T}. Emission from the secondary and bright spot are sometimes recognisable in the cited tomograms.
    
    \item EX\,Dra shows strong spiral structure in He {\sc i} $\lambda$6678, and less clearly in other lines \citep{2000A&A...356L..33J}. Emission from the secondary and the bright spot is present in outburst and quiescence \citep{1996MNRAS.278..673B}, and the H$\alpha$ tomogram in quiescence shows a second spot towards the lower right region of the map \citep{2000A&A...354..579J}. 
    
    \item U\,Gem shows emission from the secondary and strong spiral density waves in the He {\sc ii} $\lambda$4686 tomograms during outburst \citep{2001ApJ...551L..89G}. Spiral arms change in strength and opening angle with the progress of the outburst \citep{2001ApJ...551L..89G}, but remain at the same phases. Tomograms in quiescence reveal emission from a bright spot \citep{1990ApJ...364..637M}.
    
    \item WZ\,Sge shows two-armed spiral structure at slightly shifted phases than those for IP\,Peg or U\,Gem \citep{2002PASJ...54L...7B}. The pattern is strongest in He {\sc ii}, and remains visible in H$\beta$ at later stages of the outburst \citep{2004AN....325..185S}. In quiescent, tomograms show signs of a bright spot in the leading side of the disc \citep{2000MNRAS.318..429S}.
    
    \item V3885\,Sgr shows evidence of spiral structure in the Balmer tomograms, with irradiation of the secondary star  \citep{2005MNRAS.363..285H}.
    
    \item DQ\,Her exhibits spiral waves in He {\sc i} tomograms and at slightly different phases in the Balmer tomograms, but not in H$\alpha$ nor in He {\sc ii} $\lambda$4686 \citep{2010MNRAS.407.1903B}. 
    
    \item CRTS\,J1111, also SBSS 1108+574, is a Helium-rich CV at an short orbital period beyond the period minimum, that produces grazing eclipses \citep{2013MNRAS.431..372C}. He {\sc i} tomograms reveal emission from tentative spiral arms \citealt{2013MNRAS.431..372C}. H$\alpha$ also shows an emitting component that could be related to a secondary bright spot or spiral structure \citep{2013AJ....145..145L}.

    \item V406\,Vir's H$\alpha$ tomograms show several regions of enhanced emission at different epochs, being \citet{2019MNRAS.483.1080P} the most compelling evidence of spiral structure. Tomograms of other lines do not show spiral waves, but their presence could also explain the double-hump modulation in the light curve (\citealt{2019MNRAS.483.1080P}, \citealt{2010ApJ...711..389A}).
    
    \item EC21178-5417 shows a strong spiral pattern in a diversity of lines (\citealt{Khangale:2013uq}, \citealt{RuizCarmona_ec}). There are night-to-night changes in strength and location of the spiral structure \citep{RuizCarmona_ec}.

\end{itemize}

These are systems that display non-axisymmetric discs in their tomogram, but the position of the regions of enhanced emission is not at the expected phases:

\begin{itemize}

    \item V2779\,Oph shows two localized regions of enhanced emission at the bottom right and top left of the Balmer tomograms similar to bright spots \citep{2011AstL...37..845Y}.
    
    \item UX\,UMa shows convincing evidence of enhanced emission in the H$\alpha$ tomogram at the 2008 epoch. The emission regions occupy the lower quadrants of the tomograms \citep{2011MNRAS.410..963N}. There are previous reports of disc winds emission, where spiral structure was not detected in the tomograms \citep{1995ApJ...448..395B}.
    
    \item HT\,Cas shows a bright spot and extended emission roughly in antiphase with the bright spot in Balmer, He {\sc i} and Fe {\sc ii} tomograms at different epochs \citep{2016A&A...586A..10N}.
    
    \item V2051\,Oph shows spiral density waves in the majority of spectral lines during quiescence in tomograms built from 8 orbital phases only \citep{2016MNRAS.463.3290R}, which can be problematic. During outburst, spiral structure was not found, and tomograms are dominated by a bright spot and a second bright spot in the leading side of the disc \citep{2015MNRAS.447..149L}.
    
    \item CRTS\,J0359 and V1838\,Aql exhibit very similar structure, with a bright spot and a second bright spot in the leading side of the disc (\citealt{2018AJ....155..232L}, \citealt{2019MNRAS.486.2631H}).

\end{itemize}


\section{Average spectra}
\label{sec:avspectra}

\begin{figure*}
  \caption{Average spectra per night of the systems observed in this work}
  \label{fig:avspec1}
  \centering
    \includegraphics[width=0.93\textwidth]{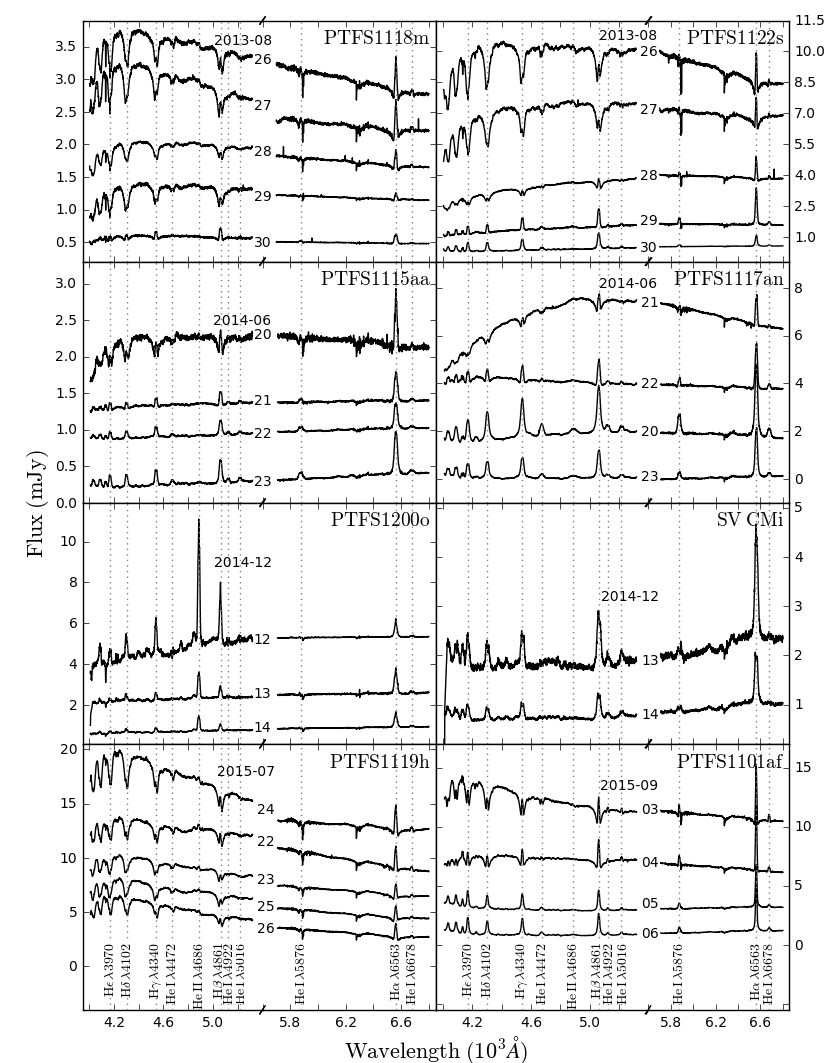}
\end{figure*}

\begin{figure*}
  \caption{Average spectra per night of the systems observed in this work (cont.)}
  \label{fig:avspec2}
  \centering
    \includegraphics[width=0.93\textwidth]{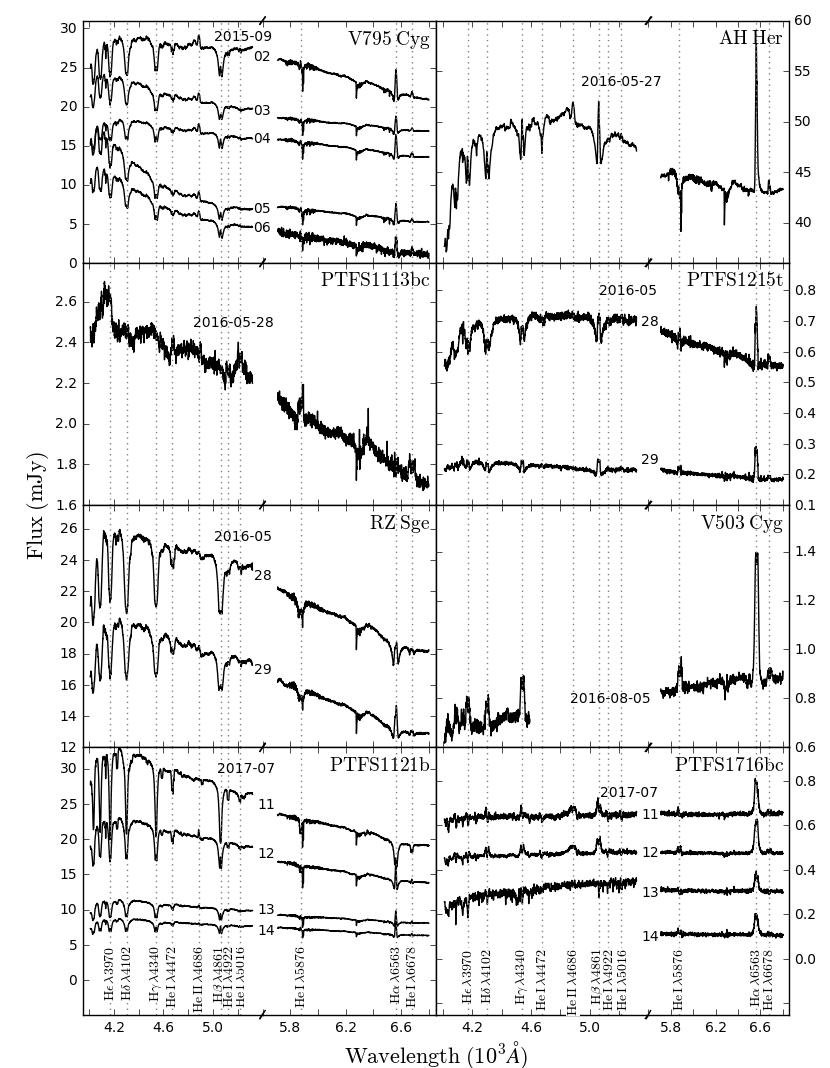}
\end{figure*}

We present an average spectrum per night for every system in Figures~\ref{fig:avspec1} and \ref{fig:avspec2}. The spectra are presented before normalization by the continuum, but we applied overall shifts to avoid overlap among the spectra at different nights. The dates are also indicated in the figures.



\section{Doppler tomography of systems exhibiting spiral waves}
\label{sec:dopplermaps}

\begin{landscape}
\begin{figure}
  \caption{Doppler tomograms and trailed spectra for PTFS\,1200o}
  \label{fig:tom_ptfs1200o}
  \centering
    \includegraphics[width=1.2\textheight]{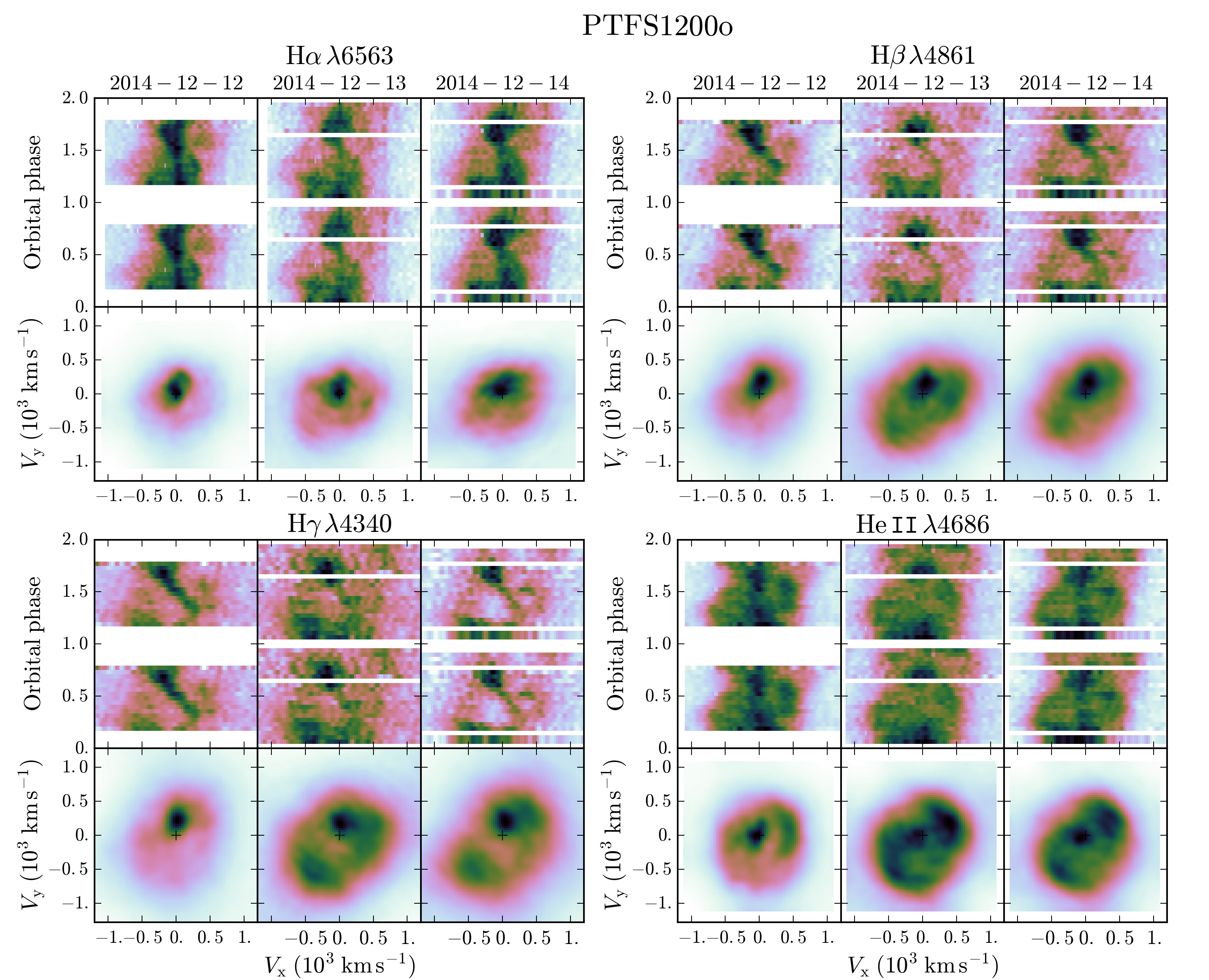}
\end{figure}
\end{landscape}

\begin{landscape}
\begin{figure}
  \caption{Doppler tomograms and trailed spectra for SV\,CMi}
  \label{fig:tom_svcmi}
  \centering
    \includegraphics[width=1\textheight]{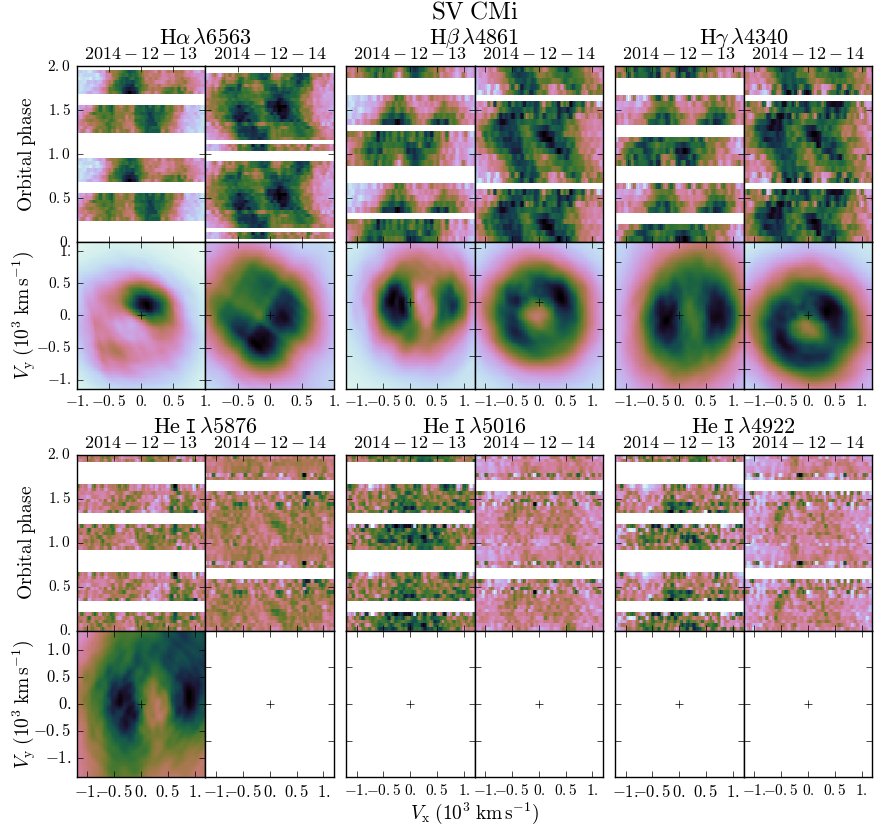}
\end{figure}
\end{landscape}

\begin{landscape}
\begin{figure}
  \caption{Doppler tomograms and trailed spectra for PTFS\,1716bc}
  \label{fig:tom_ptfs1716bc}
  \centering
    \includegraphics[width=1.2\textheight]{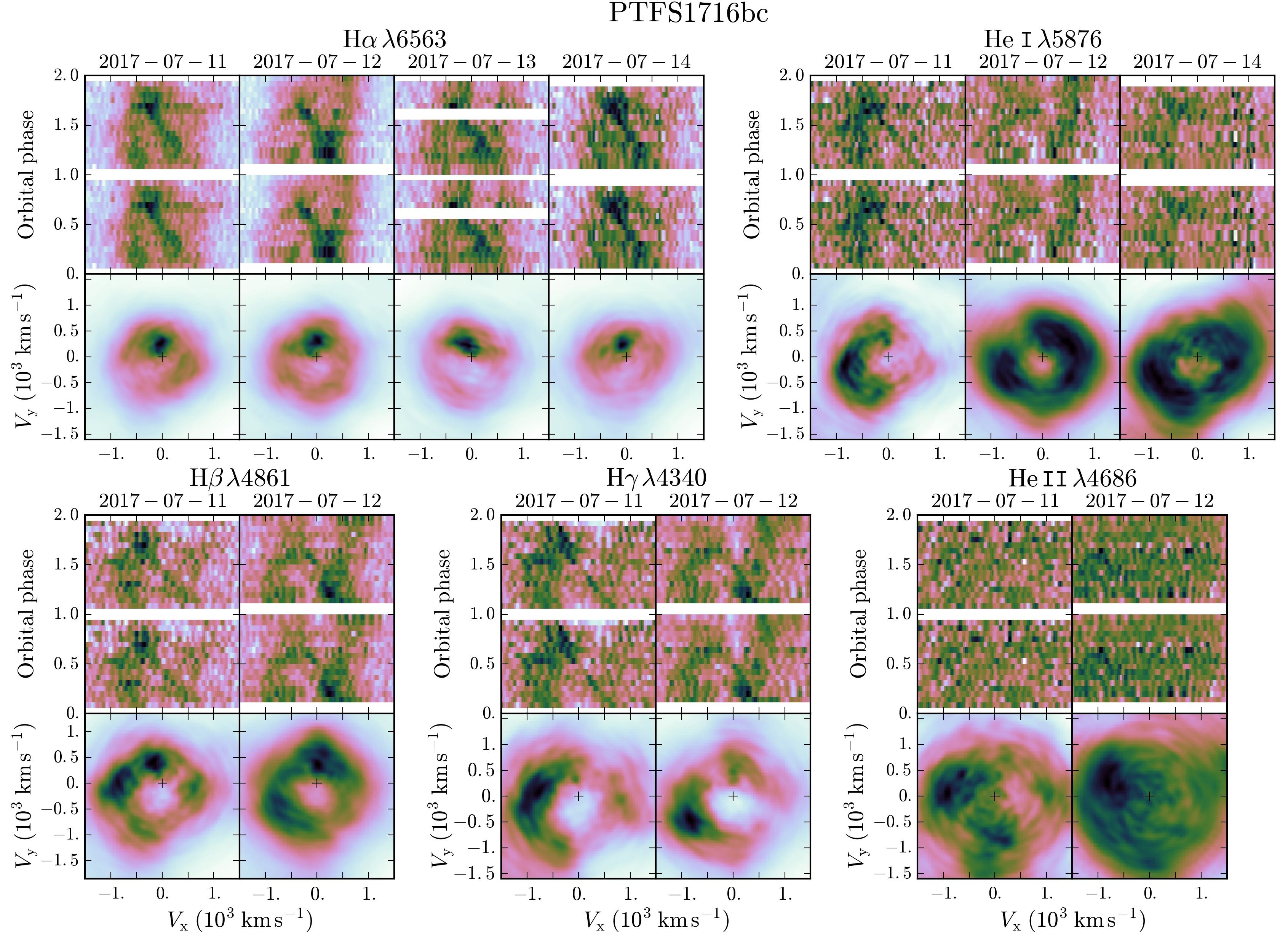}
\end{figure}
\end{landscape}

We present here the complete tomographic study of the systems that show spiral waves in their tomograms, represented in Figures~\ref{fig:tom_ptfs1200o}--\ref{fig:tom_ptfs1716bc}, including weaker lines for which significance maps were not calculated.






\bsp	
\label{lastpage}
\end{document}